\documentclass{article}
\usepackage[utf8]{inputenc}
\usepackage[english]{babel}

\usepackage{caption}
\usepackage{verbatim}
\usepackage{amsthm}
\usepackage{amssymb}
\usepackage{amsfonts}
\usepackage{amsmath}
\usepackage{anysize}

\newtheorem{theorem}{Theorem}
\newtheorem{proposition}{Proposition}

\theoremstyle{definition}
\newtheorem{definition}{Definition}
\newtheorem{example}{Example}

\usepackage{pgf,tikz}
\usetikzlibrary{chains,fit,shapes}
\usetikzlibrary{automata,arrows}

\title{Three computational models and its equivalence}
\author{Ciro Iván García López}
\date{Octubre 2020}

\begin{document}

\maketitle

\section*{Introduction}
\addcontentsline{toc}{chapter}{Introduction} 

The idea of computable operations is as old as human beings. From its beginnings, humans face up many algorithms or routines, indeed algorithms play an important role in the civilization history and one could find many examples of computable operations or devices, such as the abacus. Every day, we need to face algorithmic tasks and execute many computable operations, but what is the precise definition of computable? What can and cannot be computed?

The study of computability has its origin in Hilbert's conference of 1900, where an adjacent question, to the ones he asked, is to give a precise description of the notion of algorithm. \cite{robic2015foundations}. In the search for a good definition arose three independent theories: 

\begin{itemize}
    \item Turing and the Turing machines.
    \item Gödel and the recursive functions.
    \item Church and the Lambda Calculus. 
\end{itemize}

We refer to these three models as the classical models of computation, each of these models has its particular richness and drawbacks. As an example consider an arithmetical function, it could be easier to show that it is computable using recursive functions. In contrast, string manipulations are a very difficult and obscure task for recursiveness. This difference between classic models could lead us to think that they are different, but it is not so.

Later there were established that the classic models of computation are equivalent \cite{kleene1936}. This fact is widely accepted by many textbooks and the proof is omitted since the proof is tedious and unreadable. We intend to fill this gap presenting the proof in a modern way, without forgetting the mathematical details.

\section{Turing Model}

It is time to introduce the first classic model, the Turing Machine. This model is in some sense a generalization of automata, where there is an unlimited amount of memory and more capability of transitions. 

This section explains the Turing Machine model, it starts from an informal definition and then gives a precise one.  

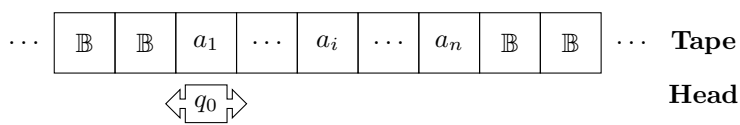
\begin{figure}
\centering
\begin{tikzpicture}
    \edef\sizetape{0.8cm}
    \tikzstyle{tmtape}=[draw,minimum size=\sizetape]
    \tikzstyle{tmhead}=[arrow box,draw,minimum size=.5cm,arrow box
    arrows={east:.25cm, west:0.25cm}]
    \begin{scope}[start chain=1 going right,node distance=-0.15mm]
        \node [on chain=1,tmtape,draw=none] {$\ldots$};
        \node [on chain=1,tmtape] { $\mathbb{B}$ };
        \node [on chain=1,tmtape] { $\mathbb{B}$ };
        \node [on chain=1,tmtape] (input) {$a_{1}$};
        \node [on chain=1,tmtape] { $\ldots$ };
        \node [on chain=1,tmtape] {$a_{i}$};
        \node [on chain=1,tmtape] { $\ldots$ };
        \node [on chain=1,tmtape] {$a_{n}$};
        \node [on chain=1,tmtape] {$\mathbb{B}$ };
        \node [on chain=1,tmtape] { $\mathbb{B}$ };
        \node [on chain=1,tmtape,draw=none] {$\ldots$};
        \node [on chain=1] (label) {\textbf{Tape}};
    \end{scope}
    \node [tmhead,yshift=-.4cm] at (input.south) (head) {$q_{0}$};
    \node [yshift=-.4cm] at (label.south) (head_label) {\textbf{Head}};
\end{tikzpicture}
\caption{The Turing Machine}
\label{fig:turing_machine}
\end{figure}

\begin{definition}
    A Turing Machine (TM) is a tuple, $M = (Q,q_{0},F,\Sigma,\Gamma,\delta)$, where:
    \begin{itemize}
        \item $Q$ is a finite set of states. 
        \item $q_{0} \in Q $ is the initial state. 
        \item $F \subseteq Q $ is the set of distinguished states called final states. 
        \item $\Sigma$ is the input alphabet. 
        \item $\Gamma$ is the tape alphabet, here $\Sigma \subseteq \Gamma$ and $\mathbb{B} \in \Gamma$. 
        \item $\delta$ are the transitions, 
        \begin{align*}
            \delta : Q \times \Gamma \to Q \times \Gamma \times \{ \leftarrow,\rightarrow \}
        \end{align*}
    \end{itemize}
\end{definition}

It is important to note that from the definition: $\mathbb{B} \not \in \Sigma$, the function $\delta$ could be a partial function, the arrow $\leftarrow$ is a movement to the left and the arrow $\rightarrow$ is a movement to the right. For example, $\delta(q,a) = (p,b,\leftarrow)$ means that when the machine is in state $q$ and reads an $a$ symbol, then changes to state $p$, writes an $b$ and moves to the left. 

When the machine starts there is an input string $w \in \Sigma^{*}$ over the tape surrounded by blanks, the head is in the first symbol of $w$ and the machine is in the initial state $q_{0}$. Figure \ref{fig:turing_machine} represents the initial stage of the machine. It is supposed that all the symbols to the left of $a_{1}$ and the right of $a_{n}$ are blank.

The following example uses the formal definition to describe a Turing machine. 

\begin{example}\label{ex:turing_formal_definition_example}
Let $\Sigma = \{0,1\}$, $\Gamma = \{0,1,X,\mathbb{B}\}$, $Q = \{q_{0},q_{1},q_{2},q_{3},q_{4},q_{5}\}$, $F=\{q_{5}\}$ and  $  \delta : Q \times \Gamma \to Q \times \Gamma \times \{ \leftarrow,\rightarrow \}$ the transition function given for the following table: 
\begin{center}
\begin{tabular}{ c  | c | c | c | c |}
     & 0 & 1 & $X$ & $\mathbb{B}$ \\
    \hline 
    $q_{0}$ & $(q_{1},X,\rightarrow)$ & & & $(\mathbb{B},\mathbb{B},\rightarrow)$ \\
    \hline 
    $q_{1}$ $(q_{1},0,\rightarrow)$  & $(q_{2},X,\rightarrow)$ & & & \\
    \hline 
    $q_{2}$ & $(q_{3},X,\leftarrow)$ & & & \\
    \hline 
    $q_{3}$ & $(q_{3},0,\leftarrow)$ & $(q_{3},1,\leftarrow)$ & $(q_{3},X,\leftarrow)$ & $(q_{4},\mathbb{B},\rightarrow)$ \\
    \hline 
    $q_{4}$ & $(q_{1},X,\rightarrow)$ & & $(q_{4},X,\rightarrow)$ & \\
    \hline 
    $q_{5}$ & & & & \\
    \hline 
\end{tabular}
\end{center}
where the empty entries mean that there is no transition defined. The input that the machine accepts are strings that look like $0^{n}1^{n}$. 
\end{example}

By this point, it is difficult to describe a Turing machine using the definition. Instead one can describe the machine using state diagrams like the ones described for automata. Thus given the transition $\delta(q,a) = (p,b,\rightarrow)$ their representation is:
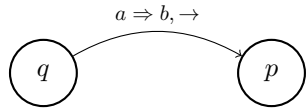
\begin{center}
\begin{tikzpicture}
	\tikzstyle{node} = [circle,thick,draw=black,minimum size=.5mm,scale=1,state]

    \node[node] (q)	at	(-3,0)	{$q$};
    \node[node] (p)	at	(0,0)	{$p$};
    
    \path[draw,->] (q) edge[bend left] node[above,scale=0.8] {$a \Rightarrow b,\rightarrow$} (p);
\end{tikzpicture}
\end{center}

The reader should be aware that complex machines have tedious descriptions, instead of giving a description using the definition or state diagram it is common to give the idea behind. Regarding informal descriptions, all the presented machines have their formal description. The below state diagram is the state diagram of example \ref{ex:turing_formal_definition_example}.

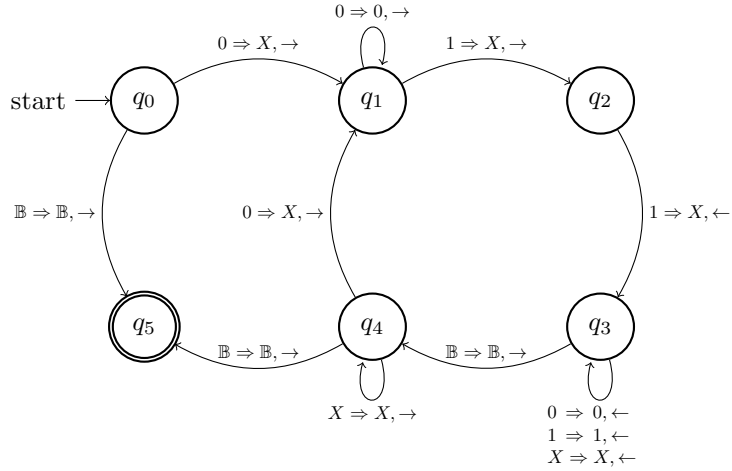
\begin{center}
\begin{tikzpicture}
	\tikzstyle{node} = [circle,thick,draw=black,minimum size=.5mm,scale=1,state]

    \node[node,initial] (node0)	at	(-3,0)	{$q_{0}$};
    \node[node] (node1)	at	(0,0)	{$q_{1}$};
    \node[node] (node2)	at	(3,0)	{$q_{2}$};
    \node[node] (node3)	at	(3,-3)	{$q_{3}$};
    \node[node] (node4)	at	(0,-3)	{$q_{4}$};
    \node[node,accepting] (node5)	at	(-3,-3)	{$q_{5}$};
    
    \path[draw,->] (node0) edge[bend left] node[above,scale=0.7] {$0 \Rightarrow X,\rightarrow$} (node1);
    \path[draw,->] (node1) edge[bend left] node[above,scale=0.7] {$1 \Rightarrow X,\rightarrow$} (node2);
    \path[draw,->] (node2) edge[bend left] node[right,scale=0.7] {$1 \Rightarrow X,\leftarrow$} (node3);
    \path[draw,->] (node3) edge[bend left] node[above,scale=0.7] {$\mathbb{B} \Rightarrow \mathbb{B},\rightarrow$} (node4);
    \path[draw,->] (node4) edge[bend left] node[left,scale=0.7] {$0 \Rightarrow X,\rightarrow$} (node1);
    \path[draw,->] (node4) edge[bend left] node[above,scale=0.7] {$\mathbb{B} \Rightarrow \mathbb{B},\rightarrow$} (node5);
    \path[draw,->] (node0) edge[bend right] node[left,scale=0.7] {$\mathbb{B} \Rightarrow \mathbb{B},\rightarrow$} (node5);
    
    \path[draw,->] (node1) edge[loop above] node[above,scale=0.7] {$0 \Rightarrow 0,\rightarrow$} (node1);
    \path[draw,->] (node3) edge[loop below] node[scale=0.7,text width=2cm] {$0 \Rightarrow 0,\leftarrow 1 \Rightarrow 1,\leftarrow  X \Rightarrow X,\leftarrow$} (node3);
    \path[draw,->] (node4) edge[loop below] node[below,scale=0.7] {$X \Rightarrow X,\rightarrow$} (node4);
\end{tikzpicture}
\end{center}

As the computation is carried out, the machine can be described by a triple containing the current state, the contents of the tape and the position of the head. This is called the configuration or snapshot of the machine. 

\begin{definition}
An expression of the form,
\begin{align*}
    a_{1}a_{2} \cdot \cdot \cdot a_{i-1}\textbf{q}a_{i} \cdot \cdot \cdot a_{n}
\end{align*}
is called a configuration or snapshot of the machine, with $a_{1},...,a_{n} \in \Gamma$, $q \in Q $. And means that the current state is $q$, the contents of the tape is $a_{1}a_{2} \cdot \cdot \cdot a_{n}$ and the head is in the position $i$. The following is the representation of a snapshot. 

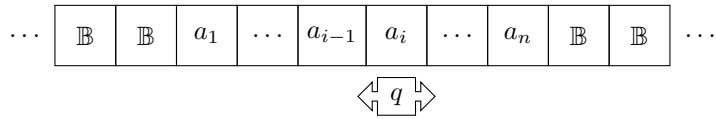
\begin{center}
\begin{tikzpicture}
    \edef\sizetape{0.8cm}
    \tikzstyle{tmtape}=[draw,minimum size=\sizetape]
    \tikzstyle{tmhead}=[arrow box,draw,minimum size=.5cm,arrow box
    arrows={east:.25cm, west:0.25cm}]
    \begin{scope}[start chain=1 going right,node distance=-0.15mm]
        \node [on chain=1,tmtape,draw=none] {$\ldots$};
        \node [on chain=1,tmtape] { $\mathbb{B}$ };
        \node [on chain=1,tmtape] { $\mathbb{B}$ };
        \node [on chain=1,tmtape] {$a_{1}$};
        \node [on chain=1,tmtape] { $\ldots$ };
        \node [on chain=1,tmtape] {$a_{i-1}$};
        \node [on chain=1,tmtape] (input) {$a_{i}$};
        \node [on chain=1,tmtape] { $\ldots$ };
        \node [on chain=1,tmtape] {$a_{n}$};
        \node [on chain=1,tmtape] {$\mathbb{B}$ };
        \node [on chain=1,tmtape] { $\mathbb{B}$ };
        \node [on chain=1,tmtape,draw=none] {$\ldots$};
    \end{scope}
    \node [tmhead,yshift=-.4cm] at (input.south) (head) {$q$};
\end{tikzpicture}
\end{center}
\end{definition}

Given a Turing Machine and an input string, it arises a sequence of configurations $c_{1},c_{2},..$. Let's suppose that the machine is in configuration $c_{i}$ and that the $\delta$ function is applied, then the machine faces the following situations: 

\begin{itemize}
    \item There is a transition defined, then the \emph{computational step} is carried out. For example if $c_{i} = u_{1}\textbf{q}au_{2}$ and $\delta(q,a) = (r,b,\rightarrow)$ then $c_{i} = u_{1}\textbf{q}au_{2} \Rightarrow u_{1}b\textbf{r}u_{2} = c_{i+1}$.
    \item There is no transition defined, then the machine stops. 
    \item The machine loops forever. This is noted by $u_{1}\textbf{q}u_{2} \overset{*}{\Rightarrow} \infty $.
\end{itemize}

The notation $\overset{*}{\Rightarrow}$ means that the machine carries out zero, one or more computational steps. In the special case when the machine reaches a state $p \in F$, it is said that the machine halts and accepts the input $w$. This motivates the following definition. 

\begin{definition}
Let $M$ be a Turing Machine, the language $L(M)$ accepted by $M$ is:
\begin{align*}
    L(M) = \{ w \in \Sigma^{*} : \mathbf{q_{0}}w \overset{*}{\Rightarrow} w_{1}\mathbf{p}w_{2},\ p \in F,\ w_{1},w_{2} \in \Gamma^{*} \}
\end{align*}
The sequence of snapshots $c_{1},c_{2},...,c_{n}$ is called a computation story. 
\end{definition}

In the literature, one can find many variants for the Turing machine. Some classical or interesting examples are: 

\begin{itemize}
    \item Non-deterministic Turing Machines.
    \item Multitape Turing Machine.
    \item Unique accepting state.
    \item Multihead Turing Machine.
\end{itemize}

Here we develop only the proofs for a few of them. The interested reader should refer to \cite{deCastroteoria,pettorossiElements,sipser2012introduction}.

\begin{proposition}
    The multitape Turing machine is equivalent to the single tape machine. 
    \begin{proof}
        The multitape Turing machine is like an ordinary machine with the difference that there are several tapes for reading and writing. It looks like: 
        
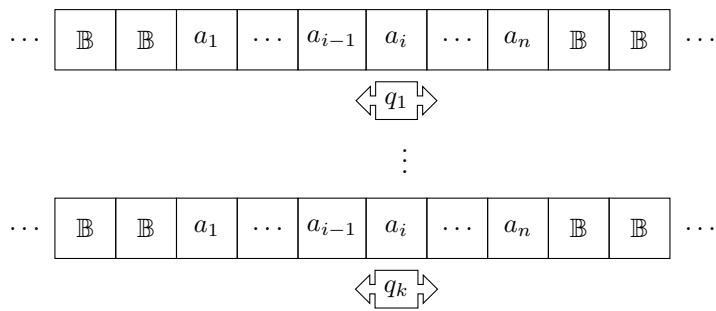
\begin{center}
            \begin{tikzpicture}
                \edef\sizetape{0.8cm}
                \tikzstyle{tmtape}=[draw,minimum size=\sizetape]
                \tikzstyle{tmhead}=[arrow box,draw,minimum size=.5cm,arrow box
                arrows={east:.25cm, west:0.25cm}]
                \begin{scope}[start chain=1 going right,node distance=-0.15mm]
                    \node [on chain=1,tmtape,draw=none] {$\ldots$};
                    \node [on chain=1,tmtape] { $\mathbb{B}$ };
                    \node [on chain=1,tmtape] { $\mathbb{B}$ };
                    \node [on chain=1,tmtape] {$a_{1}$};
                    \node [on chain=1,tmtape] { $\ldots$ };
                    \node [on chain=1,tmtape] {$a_{i-1}$};
                    \node [on chain=1,tmtape] (input) {$a_{i}$};
                    \node [on chain=1,tmtape] { $\ldots$ };
                    \node [on chain=1,tmtape] {$a_{n}$};
                    \node [on chain=1,tmtape] {$\mathbb{B}$ };
                    \node [on chain=1,tmtape] { $\mathbb{B}$ };
                    \node [on chain=1,tmtape,draw=none] {$\ldots$};
                \end{scope}
                \node [tmhead,yshift=-.4cm] at (input.south) (head) {$q_{1}$};

                \begin{scope}[xshift=5cm,yshift=-1.5cm,start chain=3 going right,node distance=-0.15mm]
                    \node [on chain=3,tmtape,draw=none] {$\vdots$};
                \end{scope}

                \begin{scope}[yshift=-2.5cm,start chain=2 going right,node distance=-0.15mm]
                    \node [on chain=2,tmtape,draw=none] {$\ldots$};
                    \node [on chain=2,tmtape] { $\mathbb{B}$ };
                    \node [on chain=2,tmtape] { $\mathbb{B}$ };
                    \node [on chain=2,tmtape] {$a_{1}$};
                    \node [on chain=2,tmtape] { $\ldots$ };
                    \node [on chain=2,tmtape] {$a_{i-1}$};
                    \node [on chain=2,tmtape] (input) {$a_{i}$};
                    \node [on chain=2,tmtape] { $\ldots$ };
                    \node [on chain=2,tmtape] {$a_{n}$};
                    \node [on chain=2,tmtape] {$\mathbb{B}$ };
                    \node [on chain=2,tmtape] { $\mathbb{B}$ };
                    \node [on chain=2,tmtape,draw=none] {$\ldots$};
                \end{scope}
                \node [tmhead,yshift=-.4cm] at (input.south) (head) {$q_{k}$};
                
            \end{tikzpicture}
        \end{center}
        
        Formally the transition function must change and looks like:
        \begin{align*}
            \delta : Q \times \Gamma^{k} \to Q \times  \Gamma^{k} \times \{\leftarrow,\rightarrow\}^{k}
        \end{align*}
        here $k$ is the number of tapes the machine has. 
        
        Let $M$ be a multitape Turing machine with $k$ tapes, the idea is to simulate $M$ by a single tape machine. Let $w_{1},...,w_{k}$ be the contents in the tapes of $M$, let $S$ be a single tape machine with input $\#w_{1}\#...\#w_{k}\#$. $S$ must know where the location of the heads is. To create the virtual heads, extend the tape alphabet and add for every $x \in \Gamma$ a dotted symbol $\dot{x}$, then $S$ knows the location of the multiple heads with the location of the dotted symbols.
        
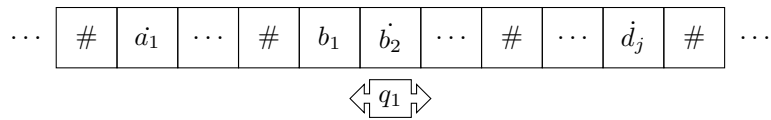
\begin{center}
        \begin{tikzpicture}
            \edef\sizetape{0.8cm}
            \tikzstyle{tmtape}=[draw,minimum size=\sizetape]
            \tikzstyle{tmhead}=[arrow box,draw,minimum size=.5cm,arrow box
            arrows={east:.25cm, west:0.25cm}]
            \begin{scope}[start chain=1 going right,node distance=-0.15mm]
                \node [on chain=1,tmtape,draw=none] {$\ldots$};
                \node [on chain=1,tmtape] { $\#$ };
                \node [on chain=1,tmtape] { $\dot{a_{1}}$ };
                \node [on chain=1,tmtape] {$ \ldots$};
                \node [on chain=1,tmtape] { $\#$ };
                \node [on chain=1,tmtape] {$b_{1}$};
                \node [on chain=1,tmtape] (input) {$\dot{b_{2}}$};
                \node [on chain=1,tmtape] { $\ldots$ };
                \node [on chain=1,tmtape] {$\#$};
                \node [on chain=1,tmtape] {$\ldots$ };
                \node [on chain=1,tmtape] { $\dot{d_{j}}$ };
                \node [on chain=1,tmtape] {$\#$};
                \node [on chain=1,tmtape,draw=none] {$\ldots$};
            \end{scope}
            \node [tmhead,yshift=-.4cm] at (input.south) (head) {$q_{1}$};
        \end{tikzpicture}
        \end{center}
    
        Then $S$ works like: 
        \begin{itemize}
            \item[] \textbf{S} $:=$ On input $w_{1},..,w_{k}$, 
            \begin{enumerate}
            \item Put in the tape of $S$ the formatted input, 
                \begin{align*}
                    \# w_{1} \# w_{2} \# ... \# w_{k}
                \end{align*}
            where the first symbol for each $w_{i}$ is dotted.    
                
            \item To simulate a move, $S$ scans its tape left to right from the first $\#$ to the $k+1$ $\#$ to determine the symbols over the virtual heads and to evaluate the transition function. Then $S$ makes a second pass to update the tape. 
            \item If $S$ reaches a $\#$ symbol, this means that $M$ has moved the corresponding head on the blank portion of the head. So $S$ must write a blank symbol and shift the tape contents according to the movement given by the transition function for this virtual head, it could move  to the right or left one place. Then continues.      
            \end{enumerate}
        \end{itemize}
        
        On the other way, every single tape Turing machine is trivially a multitape Turing machine, this completes the proof.
    \end{proof}
\end{proposition}

The next variation is the non-deterministic Turing machine. But before giving the proof for this, a few notions concerning trees would be presented and how to traverse a tree using a Turing machine. 

\begin{definition}
    A tree is a connected simple graph $G$ satisfying that $G$ does not contain a cycle. 
\end{definition}

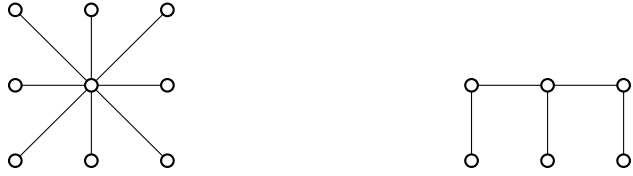
\begin{center}
\begin{tikzpicture}
    \tikzstyle{node} = [circle,thick,draw=black,minimum size=.5mm,scale=0.5]
    \node[node] (node00) at (0,0) {};
    \node[node] (node01) at (-1,-1) {};
    \node[node] (node02) at (-1,0) {};
    \node[node] (node03) at (0,-1) {};
    \node[node] (node04) at (1,1) {};
    \node[node] (node05) at (1,0) {};
    \node[node] (node06) at (0,1) {};
    \node[node] (node07) at (-1,1) {};
    \node[node] (node08) at (1,-1) {};
    
    \node[node] (node10) at (5,0) {};
    \node[node] (node11) at (6,0) {};
    \node[node] (node12) at (7,0) {};
    \node[node] (node13) at (5,-1) {};
    \node[node] (node14) at (6,-1) {};
    \node[node] (node15) at (7,-1) {};
    %\node[node] (node16) at (0,0) {};
    
    \draw (node01) -- (node00) -- (node02);
    \draw (node03) -- (node00) -- (node04);
    \draw (node05) -- (node00) -- (node06);
    \draw (node07) -- (node00) -- (node08);
    
    \draw (node13) -- (node10) -- (node11);
    \draw (node14) -- (node11) -- (node12);
    \draw (node15) -- (node12);
\end{tikzpicture}
\end{center}

Some trees have a particular element called the root. Trees with root look like: 

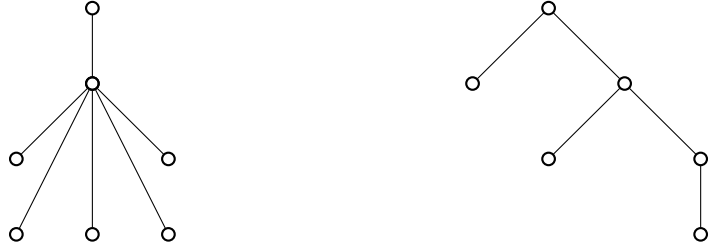
\begin{center}
\begin{tikzpicture}
    \tikzstyle{node} = [circle,thick,draw=black,minimum size=.5mm,scale=0.5]
    \node[node] (node04) at (0,2) {};
    \node[node] (node00) at (0,1) {};
    
    \node[node] (node01) at (-1,-1) {};
    \node[node] (node02) at (-1,0) {};
    \node[node] (node03) at (0,-1) {};
    
    \node[node] (node05) at (1,0) {};
    \node[node] (node06) at (0,1) {};

    \node[node] (node08) at (1,-1) {};
    
    \node[node] (node10) at (6,2) {};
    \node[node] (node11) at (7,1) {};
    \node[node] (node12) at (8,0) {};
    \node[node] (node13) at (5,1) {};
    \node[node] (node14) at (6,0) {};
    \node[node] (node15) at (8,-1) {};
    %\node[node] (node16) at (0,0) {};
    
    \draw (node01) -- (node00) -- (node02);
    \draw (node03) -- (node00) -- (node04);
    \draw (node05) -- (node00) -- (node06);
    \draw  (node00) -- (node08);
    
    \draw (node13) -- (node10) -- (node11);
    \draw (node14) -- (node11) -- (node12);
    \draw (node15) -- (node12);
\end{tikzpicture}
\end{center}

\begin{definition}
    A tree traversal is the process of visiting each node of the tree once. 
\end{definition}

\begin{proposition}
    It is possible to traverse a given tree following the depth first search (DFS) and breath first search (BFS) strategies using a Turing machine. 
    \begin{proof}
        For the proposition suppose that the given tree has at most $b$ children per node and each node operates the input in some way. Here we consider the alphabet $\Gamma_{b} = \Gamma \cup \{1,...,b\}$ and assign the address 312 for the node that is reached from the root following the third child, then the first and second for last. The addresses are strings formed with symbols $\{1,2,..,b\}$ and ordered in the following way, $s_{1} \leq s_{2}$ if $s_{2}$ is a prefix of $s_{1}$ or there is an $i$ such that $s_{1,i} \leq s_{2,i}$. 
        
        To complete the traverse, a Turing machine with three tapes is consider. In the first tape there is a read-only copy of the input, the second is the tape to operate and the last tape stores the address for the actual node. The machine description is: 
        \begin{itemize}
            \item[] \textbf{M} $:=$ On input $w$, 
            \begin{enumerate}
            \item Put $w$ in the first tape and leave the second and third tapes blank. 
            \item Copy the contents of tape $1$ in tape $2$ and put the first address in the third tape. 
            \item Use tape two to operate the input on the node reached by the given address. 
            \item Replace the string in the third tape with the following address, given by the order.
            \end{enumerate}
        \end{itemize}
        
        To traverse a tree using BFS it is sufficient to use the lexicographic order and use the same idea. 
    \end{proof}
\end{proposition}

\begin{proposition}
    The non-deterministic variant is equivalent to the deterministic machine. 
    \begin{proof}
        Again, the transition function change and looks like:
        \begin{align*}
            \delta : Q \times \Gamma \to  \mathcal{P}(Q \times  \Gamma \times \{\leftarrow,\rightarrow\}),
        \end{align*}
        and then the non-deterministic machine determines a tree, possibly infinite, that looks like:
        
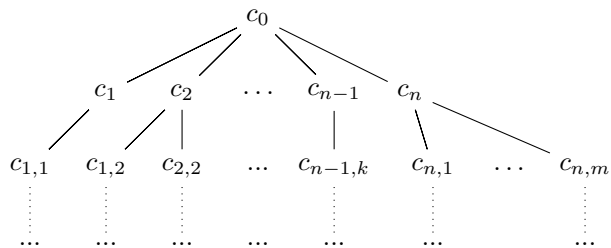
\begin{center}
        \begin{tikzpicture}
    	\node (nodo01) at (0,0) {$c_{0}$};
        
        \node (nodo11) at (-2,-1) {$c_{1}$};
        \node (nodo12) at (-1,-1) {$c_{2}$};
        \node (nodo13) at (0,-1) {$\ldots$};
        \node (nodo14) at (1,-1) {$c_{n-1}$};
        \node (nodo15) at (2,-1) {$c_{n}$};
        
        \node (nodo21) at (-3,-2) {$c_{1,1}$};
        \node (nodo22) at (-2,-2) {$c_{1,2}$};
        \node (nodo23) at (-1,-2) {$c_{2,2}$};
        \node (nodo25) at (0,-2) {...};
        \node (nodo26) at (1,-2) {$c_{n-1,k}$};
        \node (nodo27) at (2.3,-2) {$c_{n,1}$};
        \node (nodo28) at (3.3,-2) {$\ldots$};
        \node (nodo29) at (4.3,-2) {$c_{n,m}$};
        
        \node (nodo31) at (-3,-3) {...};
        \node (nodo32) at (-2,-3) {...};
        \node (nodo33) at (-1,-3) {...};
        \node (nodo35) at (0,-3)  {...};
        \node (nodo36) at (1,-3)  {...};
        \node (nodo37) at (2.3,-3) {...};
        \node (nodo38) at (4.3,-3) {...};

        \draw (nodo01) -- (nodo11) -- (nodo01) -- (nodo12) -- (nodo01) -- (nodo14) -- (nodo01) -- (nodo15) ;
        
        \draw (nodo11) -- (nodo21) -- (nodo11) -- (nodo21);
        \draw (nodo12) -- (nodo22) -- (nodo12) -- (nodo23);
        \draw (nodo14) -- (nodo26);
        \draw (nodo15) -- (nodo27) -- (nodo15) -- (nodo29);
        
        \draw [dotted] (nodo21) -- (nodo31);
        \draw [dotted] (nodo22) -- (nodo32);
        \draw [dotted] (nodo23) -- (nodo33);
        \draw [dotted] (nodo26) -- (nodo36);
        \draw [dotted] (nodo27) -- (nodo37);
        \draw [dotted] (nodo29) -- (nodo38);
        \end{tikzpicture}	
        \end{center} 
        where each $c_{i}$ is a configuration of the machine and each bifurcation is generated by a non-deterministic choice. Use the BFS strategy to traverse the tree generated by the non-deterministic machine. Here the node operates the input simulating the branches of the non-deterministic machine and halts if an accepting or rejecting state is reached. 
        
        Trivially a deterministic Turing machine is non-deterministic and the proposition is complete.
    \end{proof}
\end{proposition}

So far were presented some basic notions about the Turing models and the equivalence with some relevant variations of the model and therefore one could refer to this just as Turing machines. However, less has been said about the accepted language  by the machine. Now we develop some definitions common in the literature concerning the accepted language. 

\begin{definition}
A language $L$ is said Turing-recognizable if some Turing machine recognizes it. i.e. there is a Turing machine $M$ such that $L = L(M)$.
\end{definition}

\begin{definition}\label{def:turing_computable_set}
A language $L$ is called Turing-decidable or decidable if some Turing machine recognizes it and for every input these machine halts. 
\end{definition}

Let $L$ be a decidable language and $M$ the Turing machine that recognizes it. For every input, the machine always halts then two options are available to accept or reject, never loops. Thus these kind of machines are commonly called deciders. The following are simple results concerning decidable languages. 

\begin{proposition}
    Let $A,B$ decidable languages, then:
    \begin{itemize}
        \item $A \cup B$,
        \item $A \cap B$,
        \item $A - B $,
        \item $\bar{A}$,
        \item $A \bigtriangleup B$,
    \end{itemize}
    are decidable. 
    
    \begin{proof}
        The proposition follows from the fact that for $A$ and $B$ there are Turing machines that halts on every input. For the last, remember that $A \bigtriangleup B = (A-B) \cup (B-A)$
    \end{proof}
\end{proposition}

\begin{proposition}
    Given $A_{1},...,A_{n}$ decidable languages. Then then their union $(\bigcup_{1 \leq i \leq n} A_{i})$ and intersection $(\bigcap_{1 \leq i \leq n} A_{i})$ are decidable. 
    \begin{proof}
        Use the previous proposition and mathematical induction. 
    \end{proof}
\end{proposition}

\begin{proposition}\label{prop:turing_complementation}
    $A$ is a decidable language iff $A$ and $\bar{A}$ are Turing-recognizable. 
    \begin{proof}
        If the language is decidable then there is a Turing machine that halts on every input. Therefore one could give two machines such that recognize $A$ and $\bar{A}$ respectively. 
        
        For the second part, it is necessary to give a Turing Machine $M$ such that recognizes $A$ and halts on every input. But there are two machines $M_{1},M_{2}$ that recognizes $A$ and its complement. Consider the following machine,
        \begin{itemize}
            \item[] \textbf{M} $:=$ On input $w$, 
            \begin{enumerate}
            \item Run both $M_{1}$ and $M_{2}$ with input $w$.
            \item If $M_{1}$ accepts, accept. If $M_{2}$ accepts, reject. 
            \end{enumerate}
        \end{itemize}
        this machine always halts since $M_{1}$ or $M_{2}$ halts. 
    \end{proof}
\end{proposition}

Some textbooks call a language that is the complement of a Turing-recognizable a co- Turing-recognizable \cite{sipser2012introduction}. Let us suppose that the following language is given: 
\begin{align*}
    A_{DFA} = \{ \langle B, w \rangle : B \text{ is a deterministic finite automaton that accepts input } w\}
\end{align*}

Here $\langle B,w \rangle $ is an encoding for the deterministic finite automaton $B$ and a string $w$ that is accepted by the automaton. A classic problem related is testing the membership of any given automaton and input string, this is the same as testing whether a particular automaton accepts a given input. This problem is commonly known as the acceptance problem for finite automata. The following proposition exhibits some examples of related problems. 

\begin{proposition}
    The following are decidable languages: 
    \begin{enumerate}
        \item $A_{DFA} = \{ \langle B, w \rangle : B \text{ is a DFA that accepts input } w\}$.
        \item $A_{NDFA} = \{ \langle B, w \rangle : B \text{ is a NDFA that accepts input } w\}$.
        \item $A_{Rex} = \{ \langle R, w \rangle : B \text{ is a DFA that accepts input } w\}$.
        \item $E_{DFA} = \{ \langle B \rangle : B \text{ is a DFA and } L(B) = \empty \}$.
    \end{enumerate}
    \begin{proof}
    For the second and third parts use Kleene's theorem for automata. 
    \end{proof}
\end{proposition}

\section{Recursive Functions}

This section introduces the notion of recursive functions, it follows the general presentation in \cite{pettorossiElements} whereas the ideas follow \cite{cooper2003computability}. 

\begin{definition}\label{def:primitive_recursive_functions}
    The set of primitive recursion functions is the smallest set of functions that includes: 
    \begin{itemize}
        \item The zero functions $Z_{k}$, 
        \begin{align*}
            Z_{k} : \mathbb{N}^{k} &\to \mathbb{N} \\
            (x_{1},...,x_{k}) &\mapsto 0 
        \end{align*}
        \item The Successor function $S$, 
        \begin{align*}
            S : \mathbb{N} &\to \mathbb{N} \\
            x &\mapsto S(x) = x+1
        \end{align*}
        \item The $i$-th projection function, 
        \begin{align*}
            Pr^{k}_{i} : \mathbb{N}^{k} &\to \mathbb{N} \\
            (x_{1},...,x_{k}) &\mapsto Pr^{k}_{i}(x_{1},...,x_{k}) = x_{i}
        \end{align*}
    \end{itemize}
    and is closed by the rules:
    \begin{itemize}
        \item[$\dag$] Composition, let $g,h_{1},...h_{l}$ then,
        \begin{gather*}
            f(x_{1},...,x_{k}) = g(h_{1}(x_{1},...,x_{k}),...,h_{l}(x_{1},...,x_{k}))
        \end{gather*}    
        is primitive recursive. 
        \item[$\dag$] Primitive recursion, let $g,h$ be primitive recursive functions then,
        \begin{align*}
            f(x_{1},...,x_{k},0) &= g(x_{1},...,x_{k}) \\
            f(x_{1},...,x_{k},m+1) &= h(x_{1},...,x_{k},m,f(x_{1},...,x_{k},m)),\ \ m > 0
        \end{align*}
        is primitive recursive. 
    \end{itemize}
\end{definition}

Sometimes it is better to write $\vec{x}$ instead of $(x_{1},..,x_{k})$. 

\begin{proposition}
    The function $f$ defined by the rules of composition and primitive recursion is uniquely determined by their recursive equations.
    \begin{proof}
        The composition follows immediately from the definition. Suppose that $f,F$ are functions determined by the primitive recursive rule, then $f(\vec{x},0) = g(\vec{x}) = F(\vec{x},0)$ and by induction one could assume that $f(\vec{x},m) = F(\vec{x},m)$ and prove that $f(\vec{x},m+1) = h(\vec{x},m,f(\vec{x},m)) = h(\vec{x},m,F(\vec{x},m))  = F(\vec{x},m+1)$. 
    \end{proof}
\end{proposition}

The following are some examples of primitive recursive functions and useful results.  
 
 \begin{proposition}\label{prop:finite_sum_primitive_functions}
    The following are primitive recursive functions:
    \begin{itemize}
        \item Identity function. 
        \item Constant functions.
        \item Sum and product of two numbers.
        \item Bounded sums and products of numbers.
        \item Exponentiation.
        \item Predecessor function.
        \item sg function.
        \item Recursive difference.
        \item Absolute difference.
    \end{itemize}
    \begin{proof}
        The following are the constructions of these functions using the rules of definition \ref{def:primitive_recursive_functions}: 
        \begin{itemize}
            \item Identity function, 
            \begin{align*}
                id(x) = P_{1}^{1}(x)
            \end{align*}
            \item Constant functions, these functions are constructed one by one,
            \begin{align*}
                \bar{1}(\vec{x}) &= S(Z_{k}(\vec{x})) \\
                \bar{2}(\vec{x}) &= S(\bar{1}(\vec{x})) \\
                &\ \vdots  \\
                \bar{n+1}(\vec{x}) &= S(\bar{n}(\vec{x})) \\
                \ &\ \vdots
            \end{align*}
            \item Sum and product of two numbers,
            \begin{align*}
                Sum(m,0) &= \bar{m}(m) \\
                Sum(m,n+1) &= S(Sum(m,n)) \\
                Prod(m,0) &= Z_{1}(m) \\
                Prod(m,n+1) &= Sum(m,Prod(m,n))
            \end{align*}
            \item Bounded sums and products of numbers, construct the sum (product) of three numbers,
            \begin{align*}
                Sum_{3}(m,n,0) &= Sum(m,n) \\
                Sum_{3}(m,n,l+1) &= S(Sum_{3}(m,n,l))
            \end{align*}
            and use the same idea to construct the sum (product) of more numbers.
            \item Exponentiation, use multiplication. 
            \item Predecessor function, 
            \begin{align*}
                Pred(0) &= Z_{1}(0) \\
                Pred(n+1) &= P_{1}^{2}(n,Pred(n))
            \end{align*}
            \item sg function,
            \begin{align*}
                sg(0) &= Z_{1}(0) \\
                sg(n+1) &= \bar{1}(n+1) 
            \end{align*}
            \item Recursive difference, noted by $m\dot{-}n$,
            \begin{align*}
                m\dot{-} 0 &= m \\
                m \dot{-} (n+1) &= Pred(m\dot{-}n)
            \end{align*}
            \item Absolute difference, noted by $|m-n|$,
            \begin{align*}
                |m-n| = Sum(m\dot{-}n, n \dot{-}m) \qedhere
            \end{align*}
        \end{itemize}  
    \end{proof}
\end{proposition}

The following two propositions illustrate the way the rules are used to construct more complex functions. 

\begin{proposition}
    Given $g_{1},...,g_{n} : \mathbb{N}^{k} \to \mathbb{N}$ primitive recursive functions, then the following functions, 
    \begin{align*}
        f(\vec{x}) &= \sum_{i=1}^{n} g_{i}(\vec{x}) \\
        h(\vec{x}) &= \prod_{i=1}^{n} g_{i}(\vec{x})
    \end{align*}
    are primitive recursive. 
    \begin{proof}
        Consider the composition of the bounded sum (product) of the values given by functions. 
    \end{proof}
\end{proposition}

\begin{proposition}
    Given $g: \mathbb{N}^{k+1} \to \mathbb{N}$ primitive recursive and $n \in \mathbb{N}$, then the following functions, 
    \begin{align*}
        f(\vec{x},n) &= \sum_{i=1}^{n} g(\vec{x},i) \\
        h(\vec{x},n) &= \prod_{i=1}^{n} g(\vec{x},i)
    \end{align*}
    are primitive recursive. 
    \begin{proof}
        First, consider the $i: \mathbb{N}^{k} \to \mathbb{N}$ constant functions for $1 \leq i \leq n $, using composition the functions $g_{i}(\vec{x}) = g(\vec{x},i(\vec{x}))$ are primitive recursive. The result follows now by using the previous proposition.
    \end{proof}
\end{proposition}

By this point, with much time and effort, one could give many examples of recursive functions, but it is not clear if one could say whether a given set is primitive recursive or not. The following definition solves this question. 

\begin{definition}
    A set $S \subseteq \mathbb{N}^{k}$, is said to be primitive recursive if:
    \begin{gather*}
    \chi_{S}(\vec{x}) = \begin{cases}
    1,\ \ \text{if $ \vec{x} \in S$} \\
    0,\ \ \text{if $ \vec{x} \not \in S$}
    \end{cases}    
    \end{gather*}
    its characteristic function is primitive recursive. 
\end{definition}

Given the previous definition, it is clear that a relation $R$ is primitive recursive if the respective characteristic function is. Remember, a relation is just a subset of $\mathbb{N}^{k}$. 

\begin{example}
    The following are examples of primitive recursive relations:    
    \begin{itemize}
        \item $n < m$.
        \item $n = m$.
        \item $ n | m$.
        \item $Prime(n)$ the relation which holds iff $n$ is a prime number.
    \end{itemize}
\end{example}

Given primitive recursive relations, it is natural to ask whether the conjunction, disjunction or negation are primitive recursive. 

\begin{proposition}
    Let $P,Q$ be primitive recursive relations, then so are $\neg P $, $P \wedge Q$ and $P \vee Q $
    \begin{proof}
        Consider, $\chi_{\neg P}(\vec{x}) = 1\dot{-} \chi_{P}(\vec{x}) $, $\chi_{P \wedge Q}(\vec{x}) = \chi_{P}(\vec{x})\chi_{Q}(\vec{x})$ and $ \chi_{P\vee Q}(\vec{x}) = sg(\chi_{P}(\vec{x}) + \chi_{Q}(\vec{x}))$. 
    \end{proof}
\end{proposition}

It is usual to call the relations predicates. One important consequence of predicates is that it allows the introduction of functions defined by cases. 

\begin{proposition}
    Given $f_{1},...,f_{n}: \mathbb{N}^{k} \to \mathbb{N}$ and $R_{1},...,R_{n} \subseteq \mathbb{N}^{k}$, then the function $h$ defined by cases given by, 
    \begin{gather*}
        h(\vec{x}) = \begin{cases}
            f_{1}(\vec{x}),\ \ \text{if $R_{1}(\vec{x})$ holds}\\
            \cdots \\
            f_{n}(\vec{x}),\ \ \text{if $R_{n}(\vec{x})$ holds}\\
        \end{cases}    
    \end{gather*}
    is primitive recursive. Here it is assumed that the relations form a partition of $\mathbb{N}^{k}$, i.e. $R_{i} \cap R_{j} = \emptyset $ provided that $i \not = j$ and $ \bigcup R_{i} = \mathbb{N}^{k}$.
    \begin{proof}
        Consider the equation given by:
        \begin{gather*}
            h(\vec{x}) = \sum_{i=1}^{n} f_{i}(\vec{x}) \chi_{R_{i}}(\vec{x})
        \end{gather*}
        and then use proposition \ref{prop:finite_sum_primitive_functions}.
    \end{proof}
\end{proposition}

\begin{proposition}\label{prop:finite_quantifiers_primitive_recursive}
    Let $R \subseteq \mathbb{N}^{k}$ be a primitive recursive relation. Then the following relations are primitive recursive: 
    \begin{gather*}
        S(\vec{x},n)\ \text{iff}\ \forall y \leq n R(\vec{x},y) \\
        T(\vec{x},n)\ \text{iff}\ \exists y \leq n R(\vec{x},y). \\ 
    \end{gather*}
    \begin{proof}
        For the first statement consider the following following relation: 
        \begin{gather*}
            \prod_{i=1}^{n} R(\vec{x},i)
        \end{gather*}
        And for the second consider: 
        \begin{gather*}
            1 \dot{-} \prod_{i=1}^{n} (1\dot{-} R(\vec{x},i)) \qedhere
        \end{gather*}
    \end{proof}
\end{proposition}

The composition and primitive recursion rules, given in definition \ref{def:primitive_recursive_functions} are the basic operators for the primitive recursive set of functions. But these are not the only ones, with the definitions given so far it is possible to define another operator for the primitive recursive set of functions. 

\begin{definition}{\textbf{Bounded minimalization}}
    Given a predicate $R$, subset of $\mathbb{N}^{k}$ the bounded minimalization operator define a new function $f: \mathbb{N}^{k+1} \to \mathbb{N}$ such that given $\vec{x}\in \mathbb{N}^{k}$ and $n \in \mathbb{N}$ then, 
    \begin{align*}
        f(\vec{x},n) = \begin{cases}
        y,\ \ \text{the minimum $y \leq n $ such that $R(\vec{x},y)$ holds, if such $y$ exists} \\ 
        0,\ \ \text{otherwise}
        \end{cases}
    \end{align*}
    It is usually noted by $f(\vec{x},n) = \mu y \leq n R(\vec{x},y)$. 
\end{definition}

\begin{proposition}
    The primitive recursive functions are closed by bounded minimalization. 
    \begin{proof}
        The idea is to show that $f$ could be constructed by using the definition of primitive recursive function. Let $R$ be the relation and consider: 
        \begin{align*}
            f(\vec{x},0) &= 0 \\
            f(\vec{x},y+1) &= \begin{cases}
                f(\vec{x},y),\ \ &\text{if $\exists z\leq y R(\vec{x},z)$} \\
                y+1,\ \ &\text{if $R(\vec{x},y+1)$ and $\forall z\leq y \neg R(\vec{x},z)$ }\\
                0,\ \ &\text{otherwise}
            \end{cases}
        \end{align*}
        Here it is necessary to use proposition \ref{prop:finite_quantifiers_primitive_recursive} and the definition by cases. 
    \end{proof} 
\end{proposition}

With the given definitions one could construct many examples of primitive recursive functions but there are other examples of non-primitive recursive functions. 

\begin{example}\label{ex:ackerman_function}
    The Ackerman function defined by, 
    \begin{align*}
        A(m,0) &= m+1 \\
        A(0,n+1) &= A(1,n) \\
        A(m+1,n+1) &= A(A(m,n+1),n)
    \end{align*}
    is not primitive recursive. 
    \begin{proof}
        The idea of the proof is to show that for every primitive recursive function $f: \mathbb{N}^{k} \to \mathbb{N}$ there is a $n$ such that $f(\vec{x}) < A(x,n)$ for each $\vec{x} \in \mathbb{N}^{k}$ with $x = max\{\vec{x}\}$. Consider the following properties of the Ackermann function:
        \begin{itemize}
            \item $A(x,n) < A(x+1,n)$,
            \item $A(x,n) < A(x,n+1)$,
            \item $A(x+1,n) \leq A(x,n+1)$,
        \end{itemize}
        these properties can be establised by using mathematical induction. The proof uses the structure of primitive recursive functions: 
        
        \begin{itemize}
            \item Zero functions, $Z_{k}(\vec{x}) = 0 <  x+1 = A(x,0)$.
            \item Successor $S(x) = x+1 = A(x,0) < A(x,1)$.
            \item Projections $Pr_{i}^{k}(\vec{x}) = x_{i} < x+1 = A(x,0)$.
            
            \item Composition, suppose that $g,h_{1},...h_{l}$ are dominated by $A(x,n_{0}),A(x,n_{1}),...,A(x,n_{l})$ respectively, $h_{j}(\vec{x}) = max\{h_{1}(\vec{x}),..,h_{l}(\vec{x})\}$ and $n=max\{n_{0},...,n_{l}\}$. Then,
            \begin{align*}
                f(\vec{x}) &= g(h_{1}(\vec{x}),...,h_{l}(\vec{x})) < A(h_{j}(\vec{x}),n) < A(A(x,n),n) \\ 
                &< A(A(x,n+1),n) = A(x+1,n+1) \leq A(x,n+2)
            \end{align*}
            Thus $f$ is dominated. 
            
            \item And for the primitive recursive, suppose that $h,g$ are dominated by $A(x,n_{1}),A(x,n_{2})$ and $y=max\{\vec{x},m\}$. Let $n = max(n_{1},n_{2})$, then $f(\vec{x},0) = g(\vec{x}) < A(x,n)$. 
            
            For the recursive step, let $y_{0} = max\{\vec{x},m,f(\vec{x},m)\}$, then there are two cases. In the first case $y_{0} \not = f(\vec{x},m)$ and by the hypothesis $f(\vec{x},m+1) = h(\vec{x},m,f(\vec{x},m)) < A(y,n)$.
            
            In the second case $y_{0} = f(\vec{x},m)$. Thus, 
            \begin{align*}
                f(\vec{x},m+1) &= h(\vec{x},m,f(\vec{x},m)) < A(f(\vec{x},m),n) \\
                &< A(A(y,n),n) < A(y,n+2)
            \end{align*}
            using the induction hypothesis it follows that $f(\vec{x},m) < A(y,n)$. 
        \end{itemize}
        
        Therefore $A$ dominates the set of primitive recursive functions and as a consequence $A$ is not a primitive recursive function. 
    \end{proof}
\end{example}

From the definition of Ackerman function in example \ref{ex:ackerman_function}, it is clear that the function is recursive, it is calling itself. Yet, ¿how can these kinds of functions be constructed? For these types of constructions are required the following definitions. 

\begin{definition}
    Let $f: \mathbb{N}^{k} \to \mathbb{N} $ and $\vec{x} \in \mathbb{N}^{k}$. It is written $f(\vec{x}) \downarrow$ if $f(\vec{x})$ is defined, and $f(\vec{x}) \uparrow$ if $f(\vec{x})$ is not defined. $f$ is called total if $f(\vec{x})$ is defined for every $\vec{x} \in \mathbb{N}^{k}$, and partial otherwise.   
\end{definition}

\begin{definition}
    Let $g: \mathbb{N}^{k+1} \to \mathbb{N}$ be a function. Then the minimalization operator $\mu$ defines a new function $f: \mathbb{N}^{k} \to \mathbb{N}$ such that for $\vec{x} \in \mathbb{N}^{k}$, 
    \begin{align*}
        f(\vec{x}) = \begin{cases}
            y,\ \ \text{exists $y \in \mathbb{N}$ such that $g(\vec{x},y) = 0$ and for every $d < y$, $g(\vec{x},d)\downarrow \not = 0$} \\
            \uparrow, \ \ \text{otherwise}
        \end{cases}
    \end{align*}
\end{definition}

With the $\mu$ operator, one could define the set of partial recursive functions, hence the recursive functions. 

\begin{definition}\label{def:partial_recursive_functions}
    The class of partial recursive functions (p.r.f.) is the smallest set of partial functions such that contains the functions from definition \ref{def:primitive_recursive_functions}. And is closed by the rules from definition \ref{def:primitive_recursive_functions} and the $\mu$ operator. A total p.r.f. is called a recursive function. 
\end{definition}

\begin{definition}
    A set $A$ is recursively enumerable (r.e.) if and only if $A = \emptyset $ or there exits a recursive function $f: \mathbb{N}^{k} \to \mathbb{N}$ such that $ A = \{f(\vec{x}) : \vec{x} \in \mathbb{N}^{k}\} $. 
\end{definition}

\begin{definition}\label{def:recursive_set}
    A set $A$ is recursive if and only if there exits a total recursive function $f$ such that for every $\vec{x} \in \mathbb{N}^{k}$ if $\vec{x} \in A$ then $f(\vec{x}) = 1 $ and $f(\vec{x}) = 0$ otherwise. 
\end{definition}

\begin{proposition}\label{prop:recursive_iff_characteristic_recursive}
    A set $A$ is recursive iff $\chi_{A}$ is recursive. 
    \begin{proof}
    The proposition follows from the definitions. 
    \end{proof}
\end{proposition}

Likewise primitive recursive relations, a relation $R$ is said to be recursive if and only if its characteristic function is recursive. Note that all the propositions concerning primitive recursiveness are still valid in the recursive context.  

\begin{proposition}
    Let $A,B$ be recursive sets. Then, 
    \begin{itemize}
        \item $A \cup B$,
        \item $A \cap B$,
        \item $A - B$,
        \item $\bar{A}$,
        \item $A \bigtriangleup B$,
    \end{itemize}
    are recursive. 
    \begin{proof}
        Use proposition \ref{prop:recursive_iff_characteristic_recursive} and the facts proved concerning primitive recursive functions. For $A \bigtriangleup B $, its characteristic function can be written as $\chi_{A \bigtriangleup B}(x) = sg( \chi_{A}(x)\dot{-}\chi_{B}(x) + \chi_{B}(x)\dot{-}\chi_{A}(x) ) $.
    \end{proof}
\end{proposition}

\begin{proposition}
    Given $A_{1},...,A_{n}$ recursive sets. Then $\bigcup_{1\leq i \leq n}A_{n}$ and $\bigcap_{1\leq i \leq n}A_{n}$ are recursive.
    \begin{proof}
        Use the previous proposition and mathematical induction. 
    \end{proof}
\end{proposition}

\begin{proposition}
    A set $A$ is recursive if and only if $A$ and $\bar{A}$ are r.e.
    \begin{proof}
        That $A$ and $\bar{A}$ are r.e. follows from the fact that $\chi_{A}$ is recursive. Now suppose that $A$ and $\bar{A}$ are r.e. sets, then there are functions $f,g : \mathbb{N}^{k} \to \mathbb{N}$ such that $A = \{f(\vec{x}) : \vec{x} \in \mathbb{N}^{k}\}$ and $\bar{A} = \{g(\vec{x}) : \vec{x} \in \mathbb{N}^{k}\}$. Thus, 
        \begin{align*}
            \chi_{A}(n) = \begin{cases}
                1,\ \ &\text{If there is a $\vec{x} \in \mathbb{N}^{k}$ such that $f(\vec{x}) = n$} \\
                0,\ \ &\text{If there is a $\vec{y} \in \mathbb{N}^{k}$ such that $g(\vec{y}) = n$}
            \end{cases}
        \end{align*}
        is recursive and $A$ is a recursive set.
    \end{proof}
\end{proposition}

\begin{proposition}\label{prop:partial_recursion_parameters_order}
    Partial recursive functions are closed by permutations of parameters. 
    \begin{proof}
        Suppose that $f: \mathbb{N}^{k} \to \mathbb{N} $ is partial recursive and $\pi$ a permutation of $\{1,...,k\}$. Then, 
        \begin{align*}
            \overline{f_{\pi}}(x_{1},...,x_{k})&= f(x_{\pi(1)},...,x_{\pi(k)}) \\
            &= f(P^{k}_{\pi(1)}(x_{1},...,x_{k}),...,P^{k}_{\pi(k)}(x_{1},...,x_{k})) 
        \end{align*}
        are partial recursive. 
    \end{proof}
\end{proposition}

\section{Lambda Calculus}

This section introduces some ideas about Lambda calculus. The ideas were taken from \cite{hankin2004introduction,fernandez2009models}. The best way to introduce the $\lambda$-calculus is to use the idea used by Hankin \cite{hankin2004introduction}. Let the following grammar be given, 
\begin{gather*}
    \langle \text{wff}\rangle \rightarrow \langle \text{variable} \rangle \ |\ \lambda x.\langle \text{wff} \rangle \ |\ (\langle \text{wff}\rangle \ \langle \text{wff}\rangle )
\end{gather*}

Following this definition one could construct expressions like $\lambda x.x$, $\lambda y.(xx)$, and many more. These usually are known as well-formed formulas and the next definition formalizes this idea. 

\begin{definition}
The set of $\lambda$-terms is noted by $\Lambda$ and is the smallest set that:
\begin{itemize}
    \item $x \in \Lambda$, for every variable $x$.
    \item (Abstractions) if $M \in \Lambda$ then $(\lambda x. M) \in \Lambda$.
    \item (Applications) if $M,N \in \Lambda$ then $(MN) \in \Lambda $.
\end{itemize}
\end{definition}

The reader familiar with mathematical logic will find this presentation similar to the construction of propositional calculus. Anyone interested in more details concerning mathematical logic should refer to \cite{dalen2004logic}. 

Some of the definitions and proofs, given for $\lambda$-calculus are carried out by complexity on terms. The following proposition and definitions illustrate this idea. 

\begin{proposition}
    Every term in $\Lambda$ has balanced parentheses.
\begin{proof}
    The proposition is true for variables, there are no parentheses. Suppose that $M,N$ have balanced parentheses, then $(\lambda x.M)$ and $(MN)$ have balanced parentheses by the induction hypothesis. 
\end{proof}
\end{proposition}

\begin{definition}
    The set of bounded variables is defined inductively by the function, $BV : \Lambda \to \mathcal{P}(Var)$, 
    \begin{align*}
        BV(x) &= \emptyset, \\
        BV(\lambda x . M ) &= BV(M) \cup \{x\}, \\
        BV(MN) &=BV(M) \cup BV(N).
    \end{align*}
\end{definition}

\begin{definition}
    The set of free variables is defined inductively by the function, $FV : \Lambda \to \mathcal{P}(Var)$,
    \begin{align*}
        FV(x) &= \{x\}, \\
        FV(\lambda x . M ) &= BV(M) - \{x\}, \\
        FV(MN) &=BV(M) \cup BV(N). 
    \end{align*}
\end{definition}

It is useful to take the $\lambda$-terms that are in some sense inside another $\lambda$-term. 

\begin{definition}
    The set of sub-terms is defined inductively by the function, $FV : \Lambda \to \mathcal{P}(\Lambda)$,
    \begin{align*}
        Sub(x) &= \{x\}, \\
        Sub(\lambda x .M ) &= Sub(M) \cup \{\lambda x.M\}, \\
        Sub(MN) &= Sub(M) \cup Sub(N) \cup \{MN\}.
    \end{align*}
\end{definition}

\begin{example}
    The following are $\lambda$-terms: 
    \begin{itemize}
        \item $\lambda x.(\lambda z.(\lambda y. (xyz)w))$.
        \item $(\lambda x.y)z$.
    \end{itemize}
\end{example}

In the previous example, some expressions are difficult to read. To simplify the notation it is common to assume that abstractions associate to the right, i.e. $\lambda x.(\lambda y.M) = \lambda x.\lambda y.M = \lambda xy.M $. And that applications associate to the left, i.e. $((MN)P) = MNP$. 

\begin{definition}
    The set $C[]$ of $\lambda$-contexts is the smallest set such that:
    \begin{itemize}
        \item $x \in C[]$.
        \item $[] \in C[]$.
        \item $C_{1}[]C_{2}[]$, $\lambda xC_{1}[] \in C[]$ if $C_{1},C_{2} \in C[]$.
    \end{itemize}
\end{definition}

The contexts can be thought as $\lambda$-term with holes. 

\begin{example}\label{ex:lambda_contexts}
    The following are contexts with one hole: 
    \begin{itemize}
        \item $\lambda x. []$.
        \item $(\lambda z.[])M$.
        \item $\lambda xy.[](\lambda z.z[])$.
    \end{itemize}
\end{example}

It is common to fill the hole appearing in a context, let $C[]$ be a context then $C[M]$ represents the filling of the holes with $M$, along the present work all the holes will be filled with the same term.

\begin{example}
    In example \ref{ex:lambda_contexts} after filling the holes with $\lambda x.x$ we get: 
    \begin{itemize}
        \item $\lambda x.(\lambda x.x)$.
        \item $(\lambda z.(\lambda x.x))M$.
        \item $\lambda xy.(\lambda x.x)(\lambda z.z(\lambda x.x))$.
    \end{itemize}
\end{example}

By this point, one could say that two terms are identical if they are equal character by character, but there is not a clear notion of equality for $\lambda$-terms. The following rules give the theory of equality \cite{hankin2004introduction} for $\lambda$-terms. 

    \begin{gather*} M = M \end{gather*}
     
    \begin{gather*} \frac{M = N }{N = M } \end{gather*}
    \begin{gather*} \frac{M = N \ \ N = L}{M = L}  \end{gather*}
    \begin{gather*} \frac{M = N }{MZ = NZ} \end{gather*}
    \begin{gather*} \frac{ M = N }{ZM = ZN } \end{gather*}
    \begin{align}
        \frac{ M = N }{\lambda x.M = \lambda  x. N }& \tag{$ \xi $}
    \end{align}
    \begin{align}\label{def:function_application_rule}
        (\lambda x.M)N &= M[x := N ] \tag{$\beta$} \\
    \end{align}

The \ref{def:function_application_rule}-rule is known as function application. One could compare this rule with the procedure calls in programming languages, in which $M[x:=N]$ is read as \emph{replace all free occurrences of $x$ in $M$ by $N$}.

\begin{example}
    Let $ (\lambda xy.yx)z $ be a $\lambda$-term, using the \ref{def:function_application_rule} rule one obtains $\lambda y.yz$. Yet, consider $ (\lambda xy.yx)y $, using the \ref{def:function_application_rule} rule one obtains $\lambda y.yy$. 
\end{example}

The difference between the first and second results, in the previous example, is that the term $N$ in the second example has a free occurrence of the variable $y$ and after the substitution is no longer free and this changes the meaning of the term. This is known as \emph{variable capture}. The following definitions try to resolve the variable capture problem. 

\begin{definition}
    $M^{\prime}$ is produced from $M$ by a change of bound variables if $M = C[\lambda x.N]$ and $M^{\prime} = C[\lambda y.(N[x := y])]$ where $y$ does not occur at all in $N$ and $C[]$ is a context with one hole.
\end{definition}

\begin{definition}
    $M$ is $\alpha$-congruent to $N$ if $N$ results from $M$ by a series of change of bound variable. It is written $M =_{\alpha} N$.
\end{definition}

\begin{example}
    The following examples illustrate the concept of $\alpha$-congruent:   
    \begin{itemize}
        \item $(\lambda x . xy)(\lambda z. zy) =_{\alpha} (\lambda w.wy )(\lambda z.zy )$.
        \item $\lambda xy . xxy =_{\alpha} \lambda yx . yyx$.
    \end{itemize}
\end{example}

\begin{proposition}
    The $\alpha$-congruent relation is an equivalence relation over the $\lambda$-terms.
    \begin{proof}
        The relation is reflexive and symmetric by the definition. For the transitivity suppose that a series of changes of bound variables are applied to $M$ and $N$ is obtained, then apply the series of changes of bound variables to obtain $L$. 
    \end{proof}
\end{proposition}

One could thinks of the $\alpha$-congruent relation using the following rules by complexity on $\lambda$-terms:

\begin{itemize}
    \item $M =_{\alpha} N$ if $M = N = x$.
    \item $M =_{\alpha} N$ if $M = M_{1}M_{2}$ and $N=N_{1}N_{2}$ and $ M_{1} =_{\alpha} N_{1}$, $ M_{2} =_{\alpha} N_{2}$.
    \item $M =_{\alpha} N$ if $M = \lambda x.M_{1}$ and $N = \lambda x.N_{1}$ and $M_{1} =_{\alpha} N_{1}$.
    \item $M =_{\alpha} N$ if $M = \lambda x.M_{1}$ and $N = \lambda y.N_{1}$ and there is a variable $z$ that not occurs in $M_{1},N_{1}$ and $M_{1}[x := z] =_{\alpha} N_{1}[y := z]$.
\end{itemize}

Therefore all the $\lambda$-terms are considered as representative of its equivalence class. And one can solve the variable capture problem defining the substitution strategy as follows: 

\begin{itemize}
    \item $x[x := N] = N$.
    \item $y[x := N] = y$.
    \item $(M_{1}M_{2})[x := N] = (M_{1}[x := N])(M_{2}[x := N])$.
    \item $(\lambda y.M )[x := N] = \lambda y .(M[x := N])$ with $y$ appearing free in $N$. 
\end{itemize}
here the term $\lambda y.M$ is an appropriate representative of the $\alpha$-equivalence class. 

\begin{theorem}{\textbf{Fixed Point Theorem.}}
    For every $F \in \Lambda$, there exists $X \in \Lambda$ such that $FX = X$.
    \begin{proof}
        Let $W = \lambda x.F(xx)$ and $X = WW$. Then, 
        \begin{gather*}
            X = (\lambda x.F(xx) )W = F(WW) = FX. \qedhere
        \end{gather*}
    \end{proof}
\end{theorem}

Now we present the Church numerals, they are the representation of the natural numbers using the $\lambda$-calculus. 

\begin{definition}
    The Church numerals are defined as follows: 
    \begin{align*}
        0 &= \lambda xy.y. \\
        1 &= \lambda xy.xy. \\
        2 &= \lambda xy.x(xy). \\
        \ \ &\ \vdots \\
        n &= \lambda xy.x(...(x(xy))). \\
        \ \ &\ \vdots \\
    \end{align*}
\end{definition}

\begin{definition}
    A combinator is a closed pure term, i.e. a term containing neither free variables nor atomic constants.
\end{definition}
        
\begin{example}
    The following are some classic examples of combinators. Consider the following terms: 
    \begin{align*}
        \mathbf{K} &\equiv \lambda xy.x. \\
        \mathbf{I} &\equiv \lambda x.x.
    \end{align*}
    All of them are combinators. 
\end{example}

\begin{definition}
    The following abbreviations shall be used: 
    \begin{align*}
        X^{m}Y &\equiv X(X(...(XY)...))\ \ \text{where there are $m$ $X$'s and $m \geq 1 $}. \\ 
        X^{0}Y &\equiv Y.      
    \end{align*}
\end{definition}

The next definitions give the notion of a $\lambda$ computable function or set. 

\begin{definition}
    Let $f: \mathbb{N}^{k} \to \mathbb{N}$ partial function. A term $F$ is said to represent $f$ if and only if for all $x_{1},...,x_{n} \in \mathbb{N}$: 
    \begin{itemize}
        \item $f(x_{1},..,x_{n}) = p $ then $F\overline{x_{1}}...\overline{x_{n}} =_{\beta} \overline{p}$.
        \item $f(x_{1},..,x_{n}) \uparrow $ then $F\overline{x_{1}}...\overline{x_{n}} $ has no nf. 
    \end{itemize}
    If so, it is said that $f$ is $\lambda$-definable. 
\end{definition}

\begin{definition}\label{def:lambda_definible_set}
    Let $A \subseteq \mathbb{N}^{k}$. Then the set is said to be $\lambda$-definable if there is term $F$ such that $F\overline{x_{1}}...\overline{x_{k}} =_{\beta} \overline{1}$ if $(x_{1},...,x_{k}) \in A$ and $F\overline{x_{1}}...\overline{x_{k}} =_{\beta} \overline{0}$ if $(x_{1},...,x_{k}) \not \in A$. 
\end{definition}
        
The $\beta$-rule is the heart of the calculation in $\lambda$-calculus, then it is necessary to explore this concept. 

\begin{definition}
    Any term of the form, 
    \begin{gather*}
        (\lambda x.\ M)N
    \end{gather*}
    is called a $\beta$-redex (reducible expression) and the corresponding term,
    \begin{gather*}
        M[x:= N]    
    \end{gather*}
    is called a contractum. If a term $P$ contains an occurrence of $(\lambda x.M)N$ and it is replaced that occurrence by $M[x:=N]$, call the result $P^{\prime}$. Then it is said that $P$ $\beta$-contracts to $P^{\prime}$ and it is written,
    \begin{gather*}
        P \twoheadrightarrow_{1\beta} P^{\prime}.
    \end{gather*}

    If $P$ is changed to a term $Q$ by a finite (possibly zero) sequence of $\beta$-contractions it is noted,
    \begin{align*}
        P \twoheadrightarrow_{\beta} Q.
    \end{align*}
\end{definition}

From the previous definition, it is clear that for the Church numerals: 
\begin{align*}
    \overline{m}FX \twoheadrightarrow_{\beta} F^{m}X.
\end{align*}    

It is natural to ask when one could say that the $\beta$ reduction process terminates. This introduces the notion of normal form.

\begin{definition}
    A term $Q$ which no contains no $\beta$-redexes is called a $\beta$-normal form or just a $\beta$-nf. The set of all $\beta$-normal forms is called $\lambda \beta$-nf. If a term $P$ $\beta$-reduces to a term $Q$ in $\beta$-nf, then $Q$ is called a $\beta$-normal form of $P$.
\end{definition}

\begin{example}\label{ex:beta-nf}
    The Church numerals are in $\beta$-nf.
\end{example}

Example \ref{ex:beta-nf} introduces the following situation, given two reductions that reach the same normal form. And this is a good behavior expected for the $\beta$-reductions and it gives the capability to make computation within it. 

\begin{proposition}
    $FV(P) \supseteq FV(Q)$, if $P \twoheadrightarrow_{\beta} Q $.
    \begin{proof}
        This follows from the definition of $\beta$ reductions and the fact that change of bound variables does not affect free variables. 
    \end{proof}
\end{proposition}

\begin{proposition}\label{prop:substitution_beta_reduction} 
    If $P \twoheadrightarrow_{\beta} P^{\prime}$ and $Q \twoheadrightarrow_{\beta} Q^{\prime}$, then:
    \begin{gather*}
        Q[x:=P] \twoheadrightarrow_{\beta} Q^{\prime}[x:= P^{\prime}]
    \end{gather*}
    \begin{proof}
        Let suppose that no free variable of $P$ and $P^{\prime}$ occurs bounded in $Q$ and $Q^{\prime}$, if so change the terms $Q$, $Q^{\prime}$ with $\alpha$-congruent terms. Now, after the substitution $Q[x:=P]$ it is clear that one could reduce all occurrences of $P$ and obtain $Q[x:=P^{\prime}]$. And from the fact that $Q \twoheadrightarrow_{\beta} Q^{\prime}$ then $Q[x:=P] \twoheadrightarrow_{\beta} Q^{\prime}[x:= P^{\prime}]$. 
    \end{proof}
\end{proposition}

\begin{theorem}{\textbf{Church-Rosser for $\twoheadrightarrow_{\beta}$ relation.}}
    If $P\twoheadrightarrow_{\beta} M $ and $P \twoheadrightarrow_{\beta} N $, then there is a term $T$ such that: 
    \begin{gather*}
        M \twoheadrightarrow_{\beta} T \text{   and   } N \twoheadrightarrow_{\beta} T 
    \end{gather*}

    The following diagram represents the situation:     
    \begin{center}
    \begin{tikzpicture}
        \tikzstyle{node} = [circle,thick,draw=black,minimum size=.5mm,scale=0.8]
        \node[node] (node00) at (0,0) {$P$};
        \node[node] (node01) at (-2,-2) {$M$};
        \node[node] (node02) at (2,-2) {$N$};
        \node[node] (node03) at (0,-4) {$T$};
        
        \draw[->>,line width=0.4mm] (node00) -- (node01);
        \draw[->>,line width=0.4mm] (node00) -- (node02);
        \draw[->>,dotted,line width=0.4mm] (node01) -- (node03);
        \draw[->>,dotted,line width=0.4mm] (node02) -- (node03);
    \end{tikzpicture}
    \end{center}
    \begin{proof}
        Refer to \cite{hankin2004introduction,hindley2008lambda}. 
    \end{proof}
\end{theorem}

\begin{proposition}
    If $P$ has a $\beta$-nf it is unique modulo $=_{\alpha}$, i.e. if $P$ has $\beta$-nf $M$ and $N$ then $M =_{\alpha} N$. 
    \begin{proof}
        By the Church-Rosser theorem, there is a term $T$ such that $N,M \twoheadrightarrow_{\beta} T$. But this implies that $M =_{\alpha} T =_{\alpha} N$ since $M,N$ have no redexes.
    \end{proof}
\end{proposition}

The proposition \ref{prop:substitution_beta_reduction} tells us that $\beta$ reduction is a compatible relation, it is not hard to see that additionally, it is reflexive and transitive. But it fails to be symmetric, hence one consider the symmetric closure of the relation.

\begin{definition}
    We say $P$ is $\beta$-equal or $\beta$-convertible to $Q$ if and only if $Q$ can be obtained from $P$ by a finite (possibly zero) sequence of $\beta$-contractions, reversed $\beta$-contractions and changes of bound variables. That is, $P =_{\beta} Q$ iff there exits $P_{0},...,P_{n}$ ($n \geq 0$) such that: 
    \begin{align*}
        P_{i} \twoheadrightarrow_{1\beta} P_{i+1}\ \ \ &P_{i+1} \twoheadrightarrow_{1\beta} P_{i}\ \ \ P_{i} =_{\alpha} P_{i+1}  \text{  for $i \leq n-1 $} \\ 
        &P_{0} = P \ \ \ \ P_{n} = Q
    \end{align*}
\end{definition}

\begin{proposition}\label{prop:substitution_beta_reduction2} 
    If $P =_{\beta} Q$, $P =_{\alpha} P^{\prime}$, $Q =_{\alpha} Q^{\prime}$, then $P^{\prime} =_{\beta} Q^{\prime}$.
    \begin{proof}
        The fact that $P =_{\beta} Q$ guaranties a sequence $P_{1},...,P_{n}$ that verify the definition, consider $P_{0} = P^{\prime}$ and $P_{n+1} = Q^{\prime}$. Hence the sequence $P_{0},...,P_{n+1}$ agree to $\beta$-equality definition of $P^{\prime}$ and $Q^{\prime}$.  
    \end{proof}
\end{proposition}

\begin{proposition}
    If $M =_{\beta} M^{\prime}$ and $N =_{\beta} N^{\prime}$, 
    \begin{gather*}
        M[x:=N] =_{\beta} M^{\prime}[x:= N^{\prime}]
    \end{gather*}
    \begin{proof}
        The proof is similar to the given in proposition \ref{prop:substitution_beta_reduction}. 
    \end{proof}
\end{proposition}

\begin{theorem}\label{th:churchu_rosser_eq}{\textbf{Church-Rosser for $=_{\beta}$ relation.}}
    If $P =_{\beta} Q $ then there exists a term $T$ such that: 
    \begin{gather*}
        P \twoheadrightarrow_{\beta} T \text{   and   } Q \twoheadrightarrow_{\beta} T 
    \end{gather*}
    \begin{proof}
        The proof is given using induction on the number of steps $k$ used in the conversion. For the case $k=0$ it is immediately. Suppose the theorem true for $k$ steps, and let $P =_{\beta} Q$ in $k+1$ steps. Then by the induction hypothesis, there is a term $T_{n}$ such that $P \twoheadrightarrow_{\beta} T_{n} $ and $P_{n} \twoheadrightarrow_{\beta} T_{n}$. Hence, if $P_{n+1} \twoheadrightarrow_{1\beta} P_{n}$ the same $T_{n}$ works for $P$ and $P_{n+1}=Q$. If $P_{n}\twoheadrightarrow_{1\beta} P_{n+1}$, then apply the Church-Rosser version for $\twoheadrightarrow_{\beta}$ and obtain a $T$ like in the figure. 
        
        
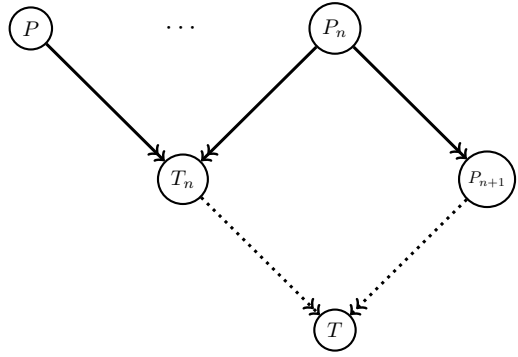
\begin{center}
        \begin{tikzpicture}
            \tikzstyle{node} = [circle,thick,draw=black,minimum size=.5mm,scale=0.8]
            \node[node] (node00) at (0,0) {$P_{n}$};
            \node[node] (node01) at (-2,-2) {$T_{n}$};
            \node[node,scale=0.8] (node02) at (2,-2) {$P_{n+1}$};
            \node[node] (node03) at (0,-4) {$T$};
            \node (node05) at (-2,0) {$\cdots$};
            
            \node[node] (node04) at (-4,0) {$P$};
            
            \draw[->>,line width=0.4mm] (node00) -- (node01);
            \draw[->>,line width=0.4mm] (node00) -- (node02);
            \draw[->>,line width=0.4mm] (node04) -- (node01);
            \draw[->>,dotted,line width=0.4mm] (node01) -- (node03);
            \draw[->>,dotted,line width=0.4mm] (node02) -- (node03);
        \end{tikzpicture}
        \end{center}
    \end{proof}
\end{theorem}

\begin{proposition}
    If $P =_{\beta} Q$ and $Q$ is in $\beta$-nf, then $P \twoheadrightarrow_{\beta} Q$.
    \begin{proof}
        By the Church-Rosser theorem $P$ and $Q$ reduce to the same $T$ but since $Q$ has no redexes $Q =_{\alpha}$. Hence $P \twoheadrightarrow_{\beta} Q$.
    \end{proof}
\end{proposition}

\begin{proposition}
    If $P=_{\beta} Q $ then either $P$, $Q$ have the same $\beta$-nf or both have no $\beta$-nf.  
    
    \begin{proof}
        It follows from the Church-Rosser theorem that if they have nf necessary is the same. Or if one does not have nf the other cannot have.
    \end{proof}
\end{proposition}

\begin{proposition}
    If $P$, $Q$ are in $\beta$-nf and $P =_{\beta} Q$ then $P =_{\alpha} Q$.
    \begin{proof}
        By the Church-Rosser theorem $P$, $Q$ has a common $T$ such that are equal modulo change of bound variables. 
    \end{proof}
\end{proposition}

\begin{proposition}
    A term is $\beta$-equal to at most one $\beta$-nf modulo changes of bound variables. 
    \begin{proof}
        Follows immediately. 
    \end{proof}
\end{proposition}

More details can be found at \cite{hindley2008lambda}, at \cite{hankin2004introduction} there is a general discussion about reductions in a more general sense.

\section{Equivalence Between Models}

Three models of computation were presented in the last section Turing Machines, Partial Recursive Functions, and Lambda Calculus. It is common to refer to these models as the classic models of computation. This section intends to prove the equivalence between the aforementioned models. The ideas for the following proofs can be found at \cite{taylor1998models,cooper2003computability,hankin2004introduction,kleene1936,hindley2008lambda}. 

\begin{theorem}\label{th:equivalence_between_classical_models}
    The classic models of computation are equivalent models.
    
    \begin{proof}
    The idea is to show how to simulate any model inside the other, for simplicity only functions $f:\mathbb{N}^{k} \to \mathbb{N}$ are considered. The proof is divided into four parts. Using the results from the previous sections for the proof it is supposed that:
    \begin{itemize}
        \item Turing machines work over the alphabet $\{0,1,\lambda\}$.
        \item The machines have a one way infinite tape with a read only blank symbol in the leftmost cell.
        \item The natural numbers are represented in the following way, $0$ is the zero and to represent $n$ it is put on the tape $n$ consecutive $1$. 
        \item After the machine reaches an acceptance state it moves the head to the first nonblank cell in the tape.
    \end{itemize}
    
    \begin{description}
    \item \textbf{p.r.f $\implies$ T.M}
        
        To show that a given partial recursive function $f: \mathbb{N} \to \mathbb{N}$ is computable by a Turing machine is sufficient to show that the Turing Machines follow the definition \ref{def:partial_recursive_functions}. 
        \begin{itemize}
            \item Let $Z$ be the Turing Machine such that: 
            \begin{itemize}
                \item[] Z $:=$ On input $x$, 
                \begin{enumerate}
                \item Erase all the content of the tape.
                \item Print a $0$.
                \end{enumerate}
            \end{itemize}
                    
            \item Let $S$ be the Turing Machine that:
            \begin{itemize}
                \item[] S $:=$ On input $x$, 
                \begin{enumerate}
                \item If the input is $0$ print a $1$, otherwise go to step 2.
                \item Go to the rightmost part of the input and append a $1$.
                \end{enumerate}
            \end{itemize}    
            
            \item Let $P_{i}^{k}$ be the Multitape machine with $k+1$ tapes such that: 
            \begin{itemize}
                \item[] $ P_{i}^{k}\ := $ On input $x_{1},...,x_{k}$, 
                \begin{enumerate}
                \item Copy the contents of tape $i$ into the tape $k+1$.
                \item Erase the contents of tapes $1$ to $k$.
                \end{enumerate}
            \end{itemize}
                
            \item Let $G,H_{1},...H_{l},$ be Turing Machines with $k+1$ tapes. Then consider the following machine: 
            \begin{itemize}
                \item[] $ F\ := $ On input $x_{1},...,x_{k}$, 
                \begin{enumerate}
                \item Run the machines $H_{1},...,H_{l}$ using the inputs $x_{1},..,x_{k}$. 
                \item Let $\dot{x}_{1},...,\dot{x}_{l}$ be the outputs of the previous executions. Then run the machine $G$ with input $\dot{x}_{1},...,\dot{x}_{l}$.
                \item Output the result from the execution of $G$.
                \end{enumerate}
            \end{itemize}    
            This machine simulates the composition of functions. 
            
            \item Let $G,H$ be Turing Machines with $k+1$ and $k+3$ tapes respectively and the order set $\mathbb{N}$. Construct the following machine with $k+4$ tapes: 
            \begin{itemize}
                \item[] $ F\ := $ On input $x_{1},...,x_{k},x$, 
                \begin{enumerate}
                \item If $x$ is zero, then run $G$ with input $x_{1},...,x_{k}$ and output the same from this execution.
                \item If $x$ is not zero then put $1$ in the $k+2$ tape and put in the $k+3$ tape the output from the execution of $G$ with input $x_{1},,,x_{k}$.
                \item Run $H$ with input $x_{1},...,x_{k}$ and the contents of the $k+2, k+3$ tape. And overwrite the tape $k+3$ with the output of this execution.
                \item If $x$ is equal to the content of the $k+2$ tape. Halts and output the content of the tape $k+3$, otherwise put the next natural number in the tape $k+2$ and repeat from step 3.
                \end{enumerate}
            \end{itemize}    
            The machine is the simulation of the primitive recursion rule. 

            \item Let $G$ be a Turing Machine with $k+2$ tapes and the order set $\mathbb{N}$. Consider: 
            \begin{itemize}
                \item[] $ F\ := $ On input $x_{1},...,x_{k}$, 
                \begin{enumerate}
                \item Select the first natural number $x$ and put it in the $k+1$ tape. 
                \item Run the machine $G$ with input $x_{1},...,x_{k},x$.
                \item If the previous execution halts and output $0$, halt and output $x$. Otherwise, replace $x$ with the next natural number and execute step 2. 
                \end{enumerate}
            \end{itemize}    
            The machine $F$ simulates the minimalization operator $\mu$.
        \end{itemize}
        Consequently, Turing Machines for the zero exists, successor and projections functions, additionally these are closed by composition, primitive recursion and minimalization. Henceforth one could simulate any partial recursive functions using Turing Machines. 
        
    \item[] \textbf{T.M $\implies$ p.r.f }
        
        Let $M= (Q,q_{0},F,\Sigma,\Gamma,\delta)$ be a given Turing machine, the idea is to construct partial recursive functions such that  one can simulate the execution of the Turing machine. Suppose that the machine computes a partial function $f: \mathbb{N} \to \mathbb{N}$.
        
        First, it is necessary to encode any string of the alphabet, the idea behind this is to use a Gödel numbering. If the alphabet has $k$ symbols, fix a bijective map between the alphabet and the set $\{0,1,...,n\}$, and use $k$-base  to encode any string. For example in the alphabet $\{\lambda,0,1\}$ fix the map $\{ \lambda \to 0,\ 0 \to 1,\ 1 \to 2 \}$, the string $\lambda 1101\lambda \lambda \lambda...$ is encode as, 
        \begin{gather*}
            2*3^{0} + 2*3^{1} + 1*3^{1} + 2*3^{1} + 0*3^{1} + 0*3^{1} + 0*3^{1} + .... 
        \end{gather*}
        note that the first blank symbol is ignored. Then the claim is that, 
        \begin{gather*}
            Enc(n):=\ \text{n is the encoding of any number in the Turing machine}
        \end{gather*}
        is a partial recursive function. It is not difficult to see that $n$ is the encoding of a number if it is either $1$ or is of the form $3^{j}-1$ for some $j \in  \mathbb{N}$. Thus is primitive using the following definition, 
        \begin{gather*}
            Enc(n) = \begin{cases}
                1,\ \ \text{if $n=1$ or there exists $1 \leq j \leq n $ such that $ n = 3^{j}-1$} \\
                0,\ \ \text{otherwise}
            \end{cases}
        \end{gather*}
        Then, tape squares are associated to natural numbers starting with $0$ for the leftmost square, 1 with the next one, 2 with the following one and so on. 
        
        
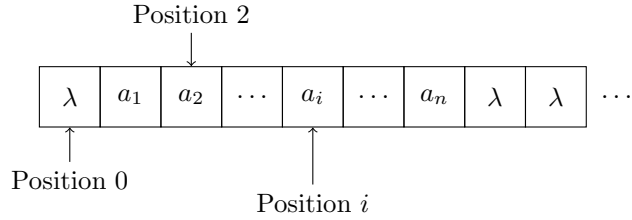
\begin{center}
        \begin{tikzpicture}
            \edef\sizetape{0.8cm}
            \tikzstyle{tmtape}=[draw,minimum size=\sizetape]
            \tikzstyle{tmhead}=[arrow box,draw,minimum size=.5cm,arrow box
            arrows={east:.25cm, west:0.25cm}]
            \begin{scope}[start chain=1 going right,node distance=-0.15mm]
                \node [on chain=1,tmtape] (1position) { $\lambda$ };
                \node [on chain=1,tmtape]  {$a_{1}$};
                \node [on chain=1,tmtape] (2position) {$a_{2}$};
                \node [on chain=1,tmtape] { $\ldots$ };
                \node [on chain=1,tmtape] (iposition) {$a_{i}$};
                \node [on chain=1,tmtape] { $\ldots$ };
                \node [on chain=1,tmtape] {$a_{n}$};
                \node [on chain=1,tmtape] {$\lambda$ };
                \node [on chain=1,tmtape] { $\lambda$ };
                \node [on chain=1,tmtape,draw=none] {$\ldots$};

            \end{scope}
            
            \node [yshift=-1cm] at (iposition.south) (i_label) {Position $i$ };
            \node [yshift=-0.7cm] at (1position.south) (1_label) {Position $0$ };
            \node [yshift=1.5cm] at (2position.south) (2_label) {Position $2$ };
            
            \draw[->] (i_label) -- (iposition);
            \draw[->] (1_label) -- (1position);
            \draw[->] (2_label) -- (2position);
        \end{tikzpicture}
        \end{center}
        
        The machine states are associated with natural numbers using the index of the state. So $q_{0}$ are with $0$, $q_{1}$ with $1$ and the other states in a similar way. It is supposed that if the machine has $r$ states, then an $r$ state it is introduced in the encoding. 
        
        Now it is time to represent the transition function as partial recursive functions. To do this, the three functions described below are necessary: 
        \begin{gather*}
            action(q,s) = \begin{cases}
            0,\ \ \text{when in state $q$ scanning the symbol $s$, the machine moves to the left} \\ 
            1,\ \ \text{when in state $q$ scanning the symbol $s$, the machine moves to the right} \\
            2,\ \ \text{otherwise}
            \end{cases}
        \end{gather*}
        \begin{gather*}
            next\_symbol(q,s) = \begin{cases}
            s^{\prime},\ \ \text{when in state $q$ scanning symbol $s$, the machine writes symbol $s^{\prime}$} \\
            s,\ \ \text{otherwise}
            \end{cases}
        \end{gather*}
        \begin{gather*}
            next\_state(q,s) = \begin{cases}
            q^{\prime},\ \ \text{when in state $q$ scanning symbol $s$, the machine enters state $q^{\prime}$} \\
            r,\ \ \text{if the machine has no instruction for the given state and symbol}
            \end{cases}
        \end{gather*}
        
        All these functions are primitive recursive since one could define them using predicates and cases. Note that: 
        \begin{itemize}
            \item If $M$ has halted, then $next\_state(q,s) = r$.
            \item If $M$ has halted, then $next\_symbol(q,s) = s$.
            \item If $M$ has halted, then $action(q,s) = 2$.
        \end{itemize}
        
        Using all the encoding introduced so far, one could represent the configuration of the machine at a time by a triple $\langle w,q,p \rangle $ where $w$ is the encoding of the string on the tape, $q$ is the actual state and $p$ is the index of the square where the head is. Now using this triple one can obtain the current symbol of the head using the following function: 
        \begin{gather*}
            current\_symbol(w,p) = (w / 3^{p\dot{-}1})\ mod\ 3
        \end{gather*}
        
        Observe that this function is partial recursive and has sense due to $div$ and $mod$ being partial recursive and that $k$ and $k^{p\dot{-}1} $ being not zero. Now it is necessary to obtain the position of the head after an action is carried out: 
        \begin{gather*}
            next\_square(w,q,p) = p \dot{-} (action(q,s) = 0) + (action(q,s)=1)
        \end{gather*}
        
        Where $s = current\_symbol(w,p)$. It is not difficult to see that the function is partial recursive and that it complies with the following rules:
        \begin{itemize}
            \item If the action is to move left, then $next\_square(w,q,p) = p \dot{-}1$.
            \item If the action is to move right, then $next\_square(w,q,p) = p + 1$.
            \item If the action to be executed is a write action, then $next\_square(w,q,p) = p$.
            \item If $M$ has halted, then $next\_square(w,q,p) = p $.
        \end{itemize}
        
        After an action, the tape contents change and it is necessary to update the value of $w$. For this, keep in mind that the tape contents are encoded using the base $3$ and if the symbol of the current position changes it is necessary to change the digit at the $3^{p\dot{-}1}$ position. Additionally, it is desired that if the machine halts the tape contents are not to change. These requirements originate the following function: 
        \begin{gather*}
            next\_tape\_contents(w,q,p) = w \dot{-} (s*3^{p\dot{-}1}) + (next\_symbol(q,s)*3^{p\dot{-}1})
        \end{gather*}

        With $s = current\_symbol(w,p)$. Now consider the function that returns the encoding of the configuration triple that results after the next instruction is executed: 
        \begin{gather*}
            next\_configuration(w,q,p) = 2^{next\_tape\_contents(w,q,p)}*3^{next\_state(q,current\_symbol(w,p))}\\
            * 5^{next_square(w,q,p)}
        \end{gather*}
        
        Note that if the machine reaches a halt state in state $q$ scanning position $p$ with tape contents $w$, then from the previous definitions $next\_configuration(w,q,p) = 2^{w}3^{r}5^{p}$. By this point, it is possible to simulate the execution of $t$ instructions using the primitive recursive rule as follows: 
        \begin{align*}
            &execute(w,0) = 2^{w}3^{0}5^{1} \\ 
            &execute(w,t+1) = next\_configuration([execute(w,t)]_{0},[execute(w,t)]_{1},[execute(w,t)]_{2})
        \end{align*}
        
        Where $[m]_{i}$ represents the extraction of the $i$ exponent in the decomposition of the number. It is necessary to show that if $M$ halts, then after $t_{h}$ steps the function $execute$ returns a constant value. 
        \begin{gather*}
            execute(w,t_{h}+1) = execute(w,t_{h}+2) = execute(w,t_{h}+3) = ... 
        \end{gather*}
        
        Note that if $M$ has halted after $t_{h}$ steps, then by definition: 
        \begin{align*}
            execute(w,t_{h}+2) &= next\_configuration([execute(w,t_{h}+1)]_{0},[execute(w,t_{h}+1)]_{1},
            \\ &\ \ \ \ \ \ \ \ \ \ \ \ \ \ \ \ [execute(w,t_{h}+1)]_{2}) \\ 
            &= next\_configuration([execute(w,t_{h}+1)]_{0},r,[execute(w,t_{h}+1)]_{2}) \\ 
            &= 2^{[execute(w,t_{h}+1)]_{0}}3^{r}5^{[execute(w,t_{h}+1)]_{2}} \\
            &= 2^{[execute(w,t_{h}+1)]_{0}}3^{[execute(w,t_{h}+1)]_{1}}5^{[execute(w,t_{h}+1)]_{2}} \\
            &= execute(w,t_{h}+1)
        \end{align*}
        
        Before defining the function computed by the machine, it is necessary to count the number of steps that the machine spent to reach a halting state. A halting state is reached when the head is at the first cell, the contents of the tape are a valid encoded number and $next\_state(q,s) = r$ when the machine is in state $q$ scanning a symbol. Here is necessary to consider the minimalization rule as follows: 
        \begin{align*}
            num\_steps(w) = \mu t [&[E]_{2} = 1 \text{ and } \\
            &Enc([E]_{0}) \text{ and } \\
            & next\_state([E]_{1},current\_symbol([E]_{0},[E]_{2})) = r ]
        \end{align*}
        
        With $E = execute(w,t)$. To sum up, if $f: \mathbb{N} \to \mathbb{N}$ is the function computed by $M$ it is seen as a partial recursive function as follows, 
        \begin{gather*}
            f(n) = \begin{cases}
                0,\ \ \text{If $[execute(N,num\_steps(N))]_{0} = 1$} \\
                j,\ \ \text{If $1\leq j$ and $\mu j ( [execute(N,num\_steps(N))]_{0} = 3^{j}-1))$ }
            \end{cases}
        \end{gather*}
        with $N$ the number encoding the natural $n$. 
        
    \item[] \textbf{p.r.f $\implies$ $\lambda$-calculus}
        
        The proof shows that the functions and rules from definition \ref{def:partial_recursive_functions} are $\lambda$-definable.
        \begin{itemize}
            \item Let $Z \equiv \lambda x.\ \overline{0}$, it follows that given any numeral $\overline{n}$, $(\lambda x. \overline{0})\overline{n}=_{\beta} \overline{0}$. 
            \item Let $S \equiv \lambda xy.\ x(uxy) $. 
            
            \item Let $P^{k}_{i} \equiv \lambda x_{1}...x_{k}.\ x_{i}$, it is clear that this term works like the projection. 
            
            \item Let suppose that $G,H_{1},...,H_{l}$ represent functions $g,h_{1},...,h_{l}$ respectively then the composition is represented by: 
            \begin{align*}
                F \equiv \lambda x_{1}...x_{k}.\ (G(h_{1}x_{1}...x_{k})...(h_{l}x_{1}...x_{k})
            \end{align*}
            
            \item Let $H$, $G$ represent $h: \mathbb{N}^{k+2} \to \mathbb{N},\ g:\mathbb{N}^{k} \to \mathbb{N}$ respectively, before giving a term for primitive recursion it is necessary to introduce some auxiliary terms. First here is consider that the primitive recursion rule is given by, 
            \begin{align*}
                f(0,\vec{x}) &= g(\vec{x}) \\
                f(m+1,\vec{x}) &= h(m,f(m,\vec{x}),\vec{x})
            \end{align*}
            
            one can use this rule as a consequence of proposition \ref{prop:partial_recursion_parameters_order}. Consider the term, 
            \begin{gather*}
                \mathbf{D} \lambda xyz.\ z(\mathbf{K}y)x
            \end{gather*}
            
            \textbf{D} is called the pairing combinator and it is such that: 
            \begin{align*}
                \mathbf{D}XY\overline{0} &\equiv (\lambda xyz.\ z(\mathbf{K}y)x)XY\overline{0} \\ 
                &\twoheadrightarrow_{\beta} \overline{0}(\mathbf{K}Y)X \\
                &\twoheadrightarrow_{\beta} (\lambda xy.\ y)(\mathbf{K}Y)X \\
                &\twoheadrightarrow_{\beta} X \\
                \\ 
                \mathbf{D}XY\overline{1} &\equiv (\lambda xyz. z(\mathbf{K}y)x)XY\overline{1} \\ 
                &\twoheadrightarrow_{\beta} \overline{1}(\mathbf{K}Y)X \\
                &\twoheadrightarrow_{\beta} (\lambda xy.\ xy)(\mathbf{K}Y)X \\
                &\twoheadrightarrow_{\beta} (\mathbf{K}Y)X \\
                &\twoheadrightarrow_{\beta} (\lambda xy.\ x)YX \\
                &\twoheadrightarrow_{\beta} Y
            \end{align*}
            
            now one can see that $\mathbf{D}XY\overline{m+1} \twoheadrightarrow_{\beta} Y $. The name given to \textbf{D} is from the fact that it gives a method of picking out the first or second member. 
            \begin{gather*}
                Q \equiv \lambda yv.\ \mathbf{D}(S(v\overline{0})) (y(v\overline{0})(v\overline{1}))
            \end{gather*}
            
            where $S$ is the successor term. One has that, 
            \begin{align*}
                QY(\mathbf{D}\overline{m}X) &\twoheadrightarrow_{\beta} \mathbf{D}(S(\mathbf{D}\overline{m}X\overline{0})) (Y( \mathbf{D}\overline{m}X\overline{0} )(\mathbf{D}\overline{m}X\overline{1} )) \\ 
                &\twoheadrightarrow_{\beta} \mathbf{D}(S\overline{m}) (Y\overline{m}X) \\ 
                &\twoheadrightarrow_{\beta} \mathbf{D}(\overline{m+1}) (Y\overline{m}X) 
            \end{align*}
            
            and for all term $X$, 
            \begin{gather*}
                (QY)^{m}(\mathbf{D}\overline{0}X) \twoheadrightarrow_{\beta} \mathbf{D}\overline{k}X_{m}
            \end{gather*}
            
            for some term $X_{m}$. Now define Bernays' \textbf{R} as follows, 
            \begin{gather*}
                \mathbf{R} \equiv \lambda xyu.\ u(Qy)(D\overline{0}x)\overline{1}
            \end{gather*}
            
            and from the construction, it is clear that,
            \begin{align*}
                \mathbf{R}XY\overline{0} &\twoheadrightarrow_{\beta} (QY)^{0}(\mathbf{D}\overline{0}X)\overline{1} \\
                &\twoheadrightarrow_{\beta} \mathbf{D}\overline{0}X\overline{1} \\
                &=_{\beta} X \\
                \\
                \mathbf{R}XY\overline{m+1} &\twoheadrightarrow_{\beta} (QY)^{m+1}(\mathbf{D}\overline{0}X)\overline{1} \\
                &\twoheadrightarrow_{\beta} (QY)((QY)^{m}(\mathbf{D}\overline{0}X))\overline{1} \\
                &\twoheadrightarrow_{\beta} QY(\mathbf{D}\overline{m}X_{m})\overline{1} \\
                &\twoheadrightarrow_{\beta} \mathbf{D}\overline{m+1}(Y\overline{m}X_{m})\overline{1} \\
                &\twoheadrightarrow_{\beta} Y\overline{m}X_{m} \\
                &=_{\beta} Y\overline{m}(\mathbf{R}XY\overline{m})
            \end{align*}
            
            Now it is possible to define the term for primitive recursion, 
            \begin{gather*}
                F \equiv \lambda u x_{1} ...x_{k}.\ \big( \mathbf{R}(Gx_{1}...x_{k})( \lambda uv\ .\ Huvx_{1}...x_{k} )u \big )
            \end{gather*}
            
            This term represents primitive recursive rule because: 
            \begin{align*}
                F \overline{0}x_{1}...x_{k} &=_{\beta} \mathbf{R}(Gx_{1}...x_{k})(\lambda uv\ .\ Huvx_{1}...x_{k})\overline{0} \\ 
                &=_{\beta} Gx_{1}...x_{k} 
            \end{align*}
            
            and 
            \begin{align*}
                F \overline{m+1}x_{1}...x_{k} &=_{\beta} \mathbf{R}(Gx_{1}...x_{k})(\lambda uv.\ Huvx_{1}...x_{k})\overline{m+1} \\
                &=_{\beta} (\lambda uv.\ Huvx_{1}...x_{k})\overline{m}(R(Gx_{1}...x_{k})(\lambda uv.\  Huvx_{1}...x_{k})\overline{m}) \\ 
                &=_{\beta} (\lambda uv.\ Huvx_{1}...x_{k})\overline{m}(F\overline{m}x_{1}...x_{k}) \\
                &=_{\beta} H\overline{m}(F\overline{m}x_{1}...x_{k})x_{1}...x_{k}
            \end{align*}
            
            \item Suppose that $G$ represents the function $g:\mathbb{N}^{k+1}\to \mathbb{N}$ the idea is to define a term $F$ that represents the $\mu$ rule over $g$. First, define: 
            \begin{align*}
                T &\equiv \lambda x.\ \mathbf{D}\overline{0}\big ( \lambda uv.\ u(x(Sv))u(Sv) \big ) \\
                P &\equiv \lambda xy.\ Tx(xy)(Tx)y
            \end{align*}
            
            with $S$ the term for the successor. For all terms $X$, $Y$, 
            \begin{align*}
                PXY &=_{\beta} TX(XY)(TX)Y \\ 
                &=_{\beta} \mathbf{D}\overline{0}( \lambda uv.\ u( X(Sv))u(Sv) ) )(XY)(TX)Y
            \end{align*}
            
            where $u,v \not \in FV(XY)$. If $XY =_{\beta} \overline{0}$, 
            \begin{align*}
                PXY &=_{\beta} \overline{0}(TX)Y \\
                &=_{\beta} Y 
            \end{align*}
            
            and if $XY =_{\beta} \overline{m+1}$ then,
            \begin{align*}
                PXY &=_{\beta} (\lambda uv.\ u(X(Sv))u(Sv))(TX)Y \\ 
                &=_{\beta} TX(X(SY))(TX)(SY) \\
                &=_{\beta} PX(SY)
            \end{align*}
            
            Henceforth it is proved that for all terms $X$, $Y$,
            \begin{align*}
                PXY &=_{\beta} Y &\ &\text{if $XY =_{\beta} \overline{0}$} \\ 
                PXY &=_{\beta} PX(SY) &\ &\text{if $XY =_{\beta} \overline{m+1}$ for some $m$} 
            \end{align*}
            
            The idea is to have a term such that simulates the function $\theta (y) = y $ if $g(x_{1},...,x_{k},y) = 0$, and moves to $\theta(y+1)$ otherwise. Defining, 
            \begin{gather*}
                H \equiv \lambda x_{1}...x_{k}.\ P(Gx_{1}...x_{k})y
            \end{gather*}
            
            it follows that for $X_{1},...X_{k},Y$,
            \begin{align*}
                HX_{1}...x_{k}Y &=_{\beta} P(GX_{1}...X_{k})Y \\
                &=_{\beta} \begin{cases}
                Y\ \ &\text{if $GX_{1}...X_{k}Y =_{\beta} \overline{0}$}\\
                HX_{1}...X_{k}(SY)\ \ &\text{if $GX_{1}...X_{k}Y =_{\beta} \overline{m+1}$ }
                \end{cases}
            \end{align*}
            
            With this, one could define,
            \begin{align*}
                J &\equiv \lambda x_{1}...x_{k}\ .\ (Hx_{1}...x_{k}\overline{0}) \\ 
                F &\equiv \lambda x_{1}...x_{k}\ .\ P(Gx_{1}...x_{k})\overline{0}\mathbf{I}(Jx_{1}...x_{k})
            \end{align*}
            
            It is clear that for all $x_{1},...,x_{k} \in \mathbb{N}$, 
            \begin{align*}
                J\overline{x_{1}}...\overline{x_{k}} =_{\beta} \overline{f(x_{1},...,x_{k})}
            \end{align*}
            
            To see that this term works for $\mu$ rule, first suppose that $x_{1},...,x_{k}$ are such that $g(x_{1},...,x_{k},d) = 0$ for some $d$ and let $y$ be the least. Then: 
            \begin{align*}
                F\overline{x_{1}}...\overline{x_{k}} &= \overline{y}\mathbf{I}(J\overline{x_{1}}...\overline{x_{k}}) \\ 
                &= \mathbf{I}^{y}(J\overline{x_{1}}...\overline{x_{k}}) \\ 
                &= J\overline{x_{1}}...\overline{x_{k}} \\ 
                &= \overline{f(x_{1},...,x_{k})}
            \end{align*}
            
            On the contrary, suppose that for $x_{1},...,x_{k}$ there is no $y$ such that $g(x_{1},...,x_{k},y) = 0$. Then it must be proved that $F\overline{x_{1}}...\overline{x_{k}}$ has no normal form. First, note that for every $y$ there is $p_{y} \geq 0$ such that:
            \begin{align*}
                g(x_{1},...,x_{k},y) = p_{y}+1
            \end{align*}
            
            Then $X\overline{y} \twoheadrightarrow_{\beta} \overline{p_{y}+1}$ if $X \equiv G\overline{x_{1}}...\overline{x_{k}}$. Because the Church numerals are in normal form for by the Church-Rosser theorem. 
            
            To prove that $F\overline{x_{1}}...\overline{x_{k}}$ has no normal form, one could find an infinite leftmost reduction of this term. Consider the following reduction: 
            \begin{align*}
                F\overline{x_{1}}...\overline{x_{k}} &\twoheadrightarrow_{\beta} PX\overline{0}\mathbf{I}G \\ &\twoheadrightarrow_{\beta} TX(X\overline{0})(TX)\overline{0}\mathbf{I}G \\ 
                &\twoheadrightarrow_{\beta} TX(\overline{p_{0}+1})(TX)\overline{0}\mathbf{I}G \\ 
                &\twoheadrightarrow_{\beta} \big( \lambda uv.\ u(X(Sv))u(Sv)  \big)(TX)\overline{0}\mathbf{I}G \\ 
                &\twoheadrightarrow_{\beta} TX(X\overline{1})(TX)\overline{1}\mathbf{I}G \\
                &\twoheadrightarrow_{\beta} \dots \\ 
                &\twoheadrightarrow_{\beta} TX(X\overline{2})(TX)\overline{2}\mathbf{I}G \\
                &\twoheadrightarrow_{\beta} \dots \text{etc.} 
            \end{align*}
            
            And this reduction is infinite. 
            
        \end{itemize}
        
        Therefore all partial recursive functions are $\lambda$-definable. And this part is complete.
    
    \item[] \textbf{ $\lambda$-calculus $\implies$ T.M }
        
        For this part, it will be constructed Turing Machines that simulates the definitions from $\lambda$-calculus.
        
        \begin{itemize}
            \item Let $V$ be the Turing Machine such that: 
                \begin{itemize}
                    \item[] V $:=$ On input $M$, 
                    \begin{enumerate}
                    \item Start reading $M$ from the leftmost character. If the character is $\lambda , ( , ) , . $ skip, otherwise copy the character in tape 2 and print out a $\#$. 
                    \item Erase all duplicate characters different from $\#$ in tape 2. 
                    \end{enumerate}
                \end{itemize}
            $V$ extracts all the variables that the term has. 
        
            \item Suppose that the alphabet includes a special character called the hole character. Let $CF$ be the Multitape Turing Machine such that: 
                \begin{itemize}
                    \item[] CF $:=$ On input $C$, $M$ tapes 1 and 2 respectively, 
                    \begin{enumerate}
                    \item Start copying the characters of the tape 1 into tape 3 from the leftmost.
                    \item Copy characters until a hole character is found, skip this and execute the next step.
                    \item Append all the content of tape 2 into tape 3. Then, return to the copy process of the previous step. 
                    \end{enumerate}
                \end{itemize}
            $CF$ simulates the process of context filling. 
                
            \item Let $CBV$ be the Multitape Turing Machine such that: 
                \begin{itemize}
                    \item[] CBV $:=$ On input $x,y,C,N$ written over tapes 1,2,3 and 4 respectively, 
                    \begin{enumerate}
                    \item Replace all occurrences of variable $x$ for $y$ in tape 4.
                    \item Append to the start of tape 3 $\lambda y.($ and append at the end a $)$ character. 
                    \item Run $CF$ with the input of tape 3 and 4.
                    \item Copy the result of $CF$ into tape 5. 
                    \end{enumerate}
                \end{itemize}
            $CBV$ simulates the process of change of bound variables. 
            
            \item Let $AE$ be the Turing Machine such that: 
                \begin{itemize}
                    \item[] AE $:=$ On input $M$, $N$ over tapes 1 and 2 respectively, 
                    \begin{enumerate}
                    \item Run the machine $V$ with input $N$. 
                    \item Print out a 0 on tape 3. 
                    \item Run a DFS search on the tree obtained from the change of bound variable, for each occurrence of an abstraction change the bound variable with each variable of the elements of $V(N)$. On each node of the search check if the is equal to $N$ if so print a 1 on tape 3.
                    \end{enumerate}
                \end{itemize}
            $AE$ check whether the terms are $\alpha$-congruent. Observe that the term $M$ is a finite sequence of symbols and that $V(N)$ is a finite set, then the search process is finite and stops.
            
            \item Let $NF$ be the Multitape Turing Machine such that: 
                \begin{itemize}
                    \item[] NF $:=$ On input $T$ over tape 1, 
                    \begin{enumerate}
                    \item Search for an occurrence of $(\lambda x.$ and match the parentheses to find the term $M$. If no occurrence of $(\lambda x.$ found, go to 3. 
                    \item If there are no more characters to the right of $(\lambda x.\ M)$, go to 3. Otherwise, write a $0$ on tape 2.
                    \item  Print out a $1$ on tape 2.
                    \end{enumerate}
                \end{itemize}
            Then, $NF$ checks if the given term is in $\beta$-normal form. 
            
            \item Let $BR$ be the Multitape Turing Machine such that: 
                \begin{itemize}
                    \item[] BR $:=$ On input $T$ over tape 1, 
                    \begin{enumerate}
                    \item Use $NF$ to check whether $T$ is in $\beta$-normal form. If so stop. 
                    \item Search for an occurrence of $(\lambda x.$ and match the parentheses to find the term $M$. 
                    \item Using parentheses match find the term $N$.
                    \item Replace all occurrences of $x$ with the hole character in $M$. 
                    \item Run $CF$ with input $M$ and $N$. 
                    \item Overwrite the content of tape 1, with the result of $CF$.
                    \item Return to the first step. 
                    \end{enumerate}
                \end{itemize}
            Then, $BR$ contracts a given term, in this process the Church-Rosser theorem \ref{th:churchu_rosser_eq} is important. Note that this machine could be modified to obtain $BR1$ that executes only one contraction.
            
        \end{itemize}
        
        Now with the help of all these Turing machines, given a $\lambda$-representation of a function $f : \mathbb{N}^{k} \to \mathbb{N}$ one could reduce $F\overline{x_{1}}...\overline{x_{k}}$ and the function is Turing computable. 
        
    \item[] \textbf{ T.M $\implies$  $\lambda$-calculus and p.r.f. $\implies$  $\lambda$-calculus}
        
        By this point, it has been proved the following: 
        
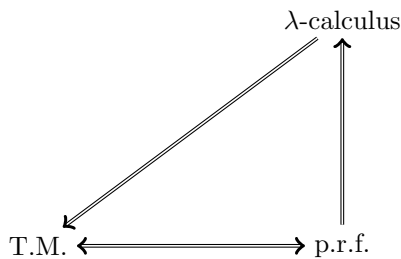
\begin{center}
        \begin{tikzpicture}
            \node (TM) at (-2,0) {T.M.};
            \node (prf) at (2,0) {p.r.f.};
            \node (lambda) at (2,3) {$\lambda$-calculus};
            
            \draw[<->,double] (TM) -- (prf);
            \draw[->,double] (lambda) -- (TM);
            \draw[->,double] (prf) -- (lambda);
        \end{tikzpicture}
        \end{center}
        Then, one could obtain the proof for this part passing through the models. 
        
        It can be proved that every function represented in $\lambda$-calculus is partial recursive. The proof uses Gödel numbering and the details can be found at \cite{kleene1936}.\qedhere
    \end{description}
    \end{proof}
\end{theorem}

The reader could make many observations from theorem \ref{th:equivalence_between_classical_models}, whereas the most important is that one could fix any of these models and develop an entire theory within it.

\newpage
\bibliography{main}
\bibliographystyle{alpha}
\nocite{*}  

\end{document}